\documentclass[iop]{emulateapj} 

\usepackage[utf8]{inputenc} 

\usepackage{graphicx} 
\usepackage{amsmath}
\usepackage{color}

\shortauthors{Balaji et al.}
\shorttitle{Tracking Starspots in Short-Period Binaries}

\begin{document}

\title{Tracking the Stellar Longitudes of Starspots in { Short-Period {\em Kepler}} Binaries}

\author{Bhaskaran Balaji\altaffilmark{1}, Bryce Croll\altaffilmark{2,3}, A. Levine\altaffilmark{2}, and S. Rappaport\altaffilmark{1} } 

\altaffiltext{1}{M.I.T. Department of Physics and Kavli
Institute for Astrophysics and Space Research, 70 Vassar St.,
Cambridge, MA, 02139; bbalaji@mit.edu; sar@mit.edu} 
\altaffiltext{2}{37-575 M.I.T. Kavli Institute for Astrophysics and Space Research, 70 Vassar St.,
Cambridge, MA, 02139; croll@space.mit.edu; aml@space.mit.edu} 
\altaffiltext{3}{5525 Olund Road, Abbotsford, B.C. Canada}

\begin{abstract}
We report on a new { method} for tracking the phases of the orbital modulations in { very short-period, near-contact, and contact} binary systems systems in order to follow starspots.  { We apply this technique} to {\em Kepler} light curves for 414 binary systems that were identified as having anticorrelated $O-C$ curves for the midtimes of the primary and secondary eclipses, or { in the case of non-eclipsing systems,  their light-curve minima}. This phase tracking { approach} extracts more  information about starspot and binary system behavior than may be easily obtained from the $O-C$ curves.  We confirm the hypothesis of Tran et al. (2013) that we can successfully follow the rotational motions of spots on the surfaces of the stars in these binaries. In $\sim$34\% of the systems, the spot rotation is retrograde as viewed in the frame rotating with the orbital motion, while $\sim$13\% show significant prograde spot rotation. The remaining systems show either little spot rotation or erratic behavior, or sometimes include intervals of both types of behavior. We discuss the possibility that the relative motions of spots are related to differential rotation of the stars.  It is clear from this study that { the motions of the starspots} in at least 50\% of { these short-period} binaries are not exactly synchronized with { the orbits}.
\end{abstract}


\section{Introduction}

Among the many targets of the {\em Kepler} mission are some 700 { short-period} binary stellar systems (with orbital periods $\lesssim 2$ days), many of them contact binaries. These systems have orbital periods that are typically less than 0.5 days, and they are exemplary candidates for timing analyses of behavior related to eclipses { or light-curve minima (in non-eclipsing systems)}. One such analysis involves the use of $O-C$ (``observed minus expected'') curves, i.e., the times of the eclipses { or light-curve minima} relative to those of a temporal reference system having a precisely constant period. Tran et al.~(2013) investigated the particular phenomenon where the $O-C$ times for the primary and secondary minima oscillate in an anticorrelated manner, found some 390 binary systems displaying anti-correlated $O-C$ curves, and reported in detail on 32 of them.  Tran et al.~(2013) concluded that a model including starspots on one of the two stars is able to explain the anticorrelated $O-C$ curves.  Their model implies or requires that the spots { move} slowly around their host stars { in the reference frame rotating with the binary}. 

In the Tran et al.~(2013) model, a spot on one of the stars produces a nearly sinusoidal variation of the apparent intensity as the binary system orbits.  The frequency of the sinusoid must be very close to that of the stellar rotation with respect to inertial space. If the star with the spot nearly corotates with the binary, then the frequency of the variation will be close to the orbital frequency, $\omega$. Since the light curves of many very short period binaries are dominated by a sinusoidal variation at the frequency $2 \omega$, the spot has the effect of slightly shifting the time of one light-curve minimum in a given direction, while shifting the time of the other light-curve minimum in the {\em opposite} sense.  Hence, the spots can induce anticorrelated behavior in the $O-C$ curves.  In this scenario, variations in the $O-C$ curves are caused by { motion of the starspots in longitude in the rotating frame of the binary}.  An additional requirement Tran et al.~(2013) presented was that the spots be visible at all times, i.e.,~not eclipsed by the companion star nor occulted by their host star.  As far as we are aware, there are no viable alternative models to explain the anticorrelated $O-C$ behavior.

\begin{figure}
\begin{center}
\includegraphics[width=\columnwidth]{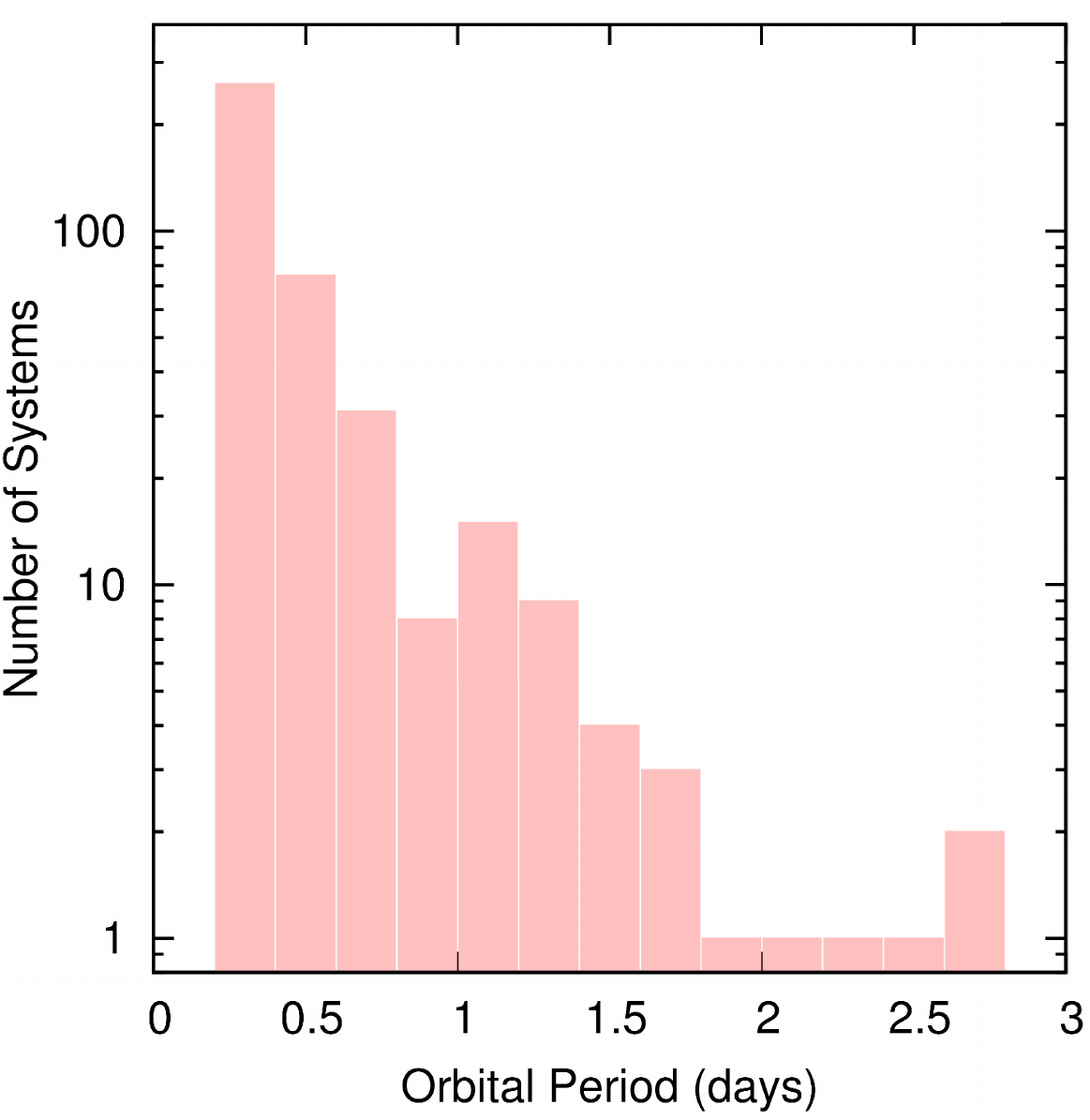}
\caption{The orbital period distribution of 412 of the 414 {\em Kepler} binaries considered in this study. To improve the resolution of the plot two systems with periods longer than 3 days have been excluded. Please note the logarithmic scaling on the y-axis.}
\label{fig:pdist}
\end{center}
\end{figure}

Despite showing that the starspot model could explain the anticorrelated $O-C$ behavior, Tran et al.~(2013) did not attempt to actually track the sinusoidal variations of the inferred starspots. In this work, we seek to support the model by explicitly tracking these variations and the inferred stellar longitudes of the spots or groups of spots for the same 414 sources that showed anticorrelated $O-C$ curves, over the duration of the {\em Kepler} mission ($\sim$4 years; Borucki et al.~2010; Caldwell et al.~2010; Batalha et al.~2013). { The orbital period distribution of the 414 binaries considered in this work is shown in Fig.~\ref{fig:pdist}.}

Section \ref{keplerData} outlines the data used in this work and the preparation thereof. Sect.~\ref{overview} presents the underlying model and demonstrates its effectiveness. Detailed analysis of a dozen systems, all exhibiting clear starspot motion is also presented. In Sect.~\ref{phasingUp}, we compare our analysis with the previously generated, but updated, $O-C$ curves with the understanding that the two methods should, if they are indeed tracking the same phenomenon, provide similar information. In Sect.~\ref{collresults}, we summarize our findings, and in Sect.~\ref{findingTriples}, we demonstrate that the results may be used to search for third bodies among the {\em Kepler} short-period binary systems.

\vspace{0.2cm}

\section{{\em Kepler} Data}
\label{keplerData}

The present study is based on {\em Kepler} long-cadence (LC) lightcurves, sets of 
sequential measurements of the flux obtained from 29.4-minute integrations. We retrieved 
the LC lightcurve files for Quarters 1 through 
17 for all of the $\sim$2600 candidates in the latest {\em Kepler} eclipsing binary catalog 
(Slawson et al.~2011; Matijevic et al.~2012) that were available at the Multimission Archive at 
STScI (MAST). We used the lightcurves made with the PDC-MAP algorithm (Stumpe et al.~2012; 
Smith et al.~2012), which is intended to remove instrumental signatures from the flux time series 
while retaining the bulk of the astrophysical variations of each target. For each quarter, the flux 
series was normalized to its median value. Then, for each target, the results from all available 
quarters were stitched together in a single flux data file.

We did not filter the data to remove long-term trends in the flux since the signal
in the binary systems of interest is due to the relatively high-frequency binary orbital modulation 
and to the nearly corotating starspots. These are the two components that are analyzed in this work.

\section{Overview of the Model}
\label{overview}

\subsection{The Orbital Terms}

The lightcurve of a very short period binary is often dominated by a sinusoid with a frequency twice that of the orbit, due to the very strong ellipsoidal light variations of the two stars that fill, or nearly fill, their Roche lobes. { To the extent that the two stars have different effective temperatures, the two lightcurve minima per orbit are expected to have different depths in eclipsing systems.  For the non-eclipsing ellipsoidal variables the cause of the difference in the depths of minimum (aside from the starspot effects) is a combination of mutual irradiation and gravity darkening.  In both cases, the differences in the depths of the two minima} can be modeled by the addition of a sinusoid at the orbital frequency. Thus, the light curves of many short-period binaries can be described adequately by the first three terms of the expression on the right-hand side of the equation:
\begin{equation}
f(t) \simeq C - E_1 \cos{\omega t} - E_2 \cos{2\omega t} - S \cos{(\omega t + \ell)}
\label{basicModel}
\end{equation}
where $C$ is a constant, $E_1$ and $E_2$ are the amplitudes that describe the orbital lightcurve of the binary, and $S$ is the starspot amplitude (discussed below). In this expression, the time $t$ is taken with respect to an origin at the primary light-curve minimum, hence the choice of negative cosine functions. The constant $C$ has a value near unity, given that it accounts for the normalized mean flux from the binary. The final term, which we explain next, describes the effect of a starspot or group of spots.

In the simplest form of the model posited in the Tran et al.~(2013) paper, the anticorrelated $O-C$ curves result from a starspot that moves slowly around one of the stars.  More generally, the $O-C$ effects may result from the combined effects of a number of spots on one or both stars.  Whether the effects are due to one or multiple spots is not of critical significance for the analyses presented below; as long as their combined effect produces a sinusoidal modulation at $\omega t$, the mathematical model in Eqn.~(\ref{basicModel}) should be sufficient.  The frequency of the motion of the spot is always close to that of the orbital motion, i.e., the star must nearly corotate with the orbit as is expected in very short period binaries. Empirically, the variations due to starspots typically show up in Fourier transforms of flux lightcurves primarily at or very close to the stellar rotation frequency, with progressively less and less power at higher harmonics (see, e.g., Rappaport et al.~2014). In the lightcurve model given by Eqn.~(\ref{basicModel}) we assume that the starspot amplitude at 2$\omega t$ is much smaller than the lightcurve amplitude $E_2$ (typically dominant in very short period binaries) and can therefore be neglected; however, this remains to be verified empirically in this work. We therefore represent the contribution of the starspot to the orbital lightcurve by the $S \cos{(\omega t + \ell)}$ term in Eqn.~(\ref{basicModel}), where $\ell$ denotes the longitude of the starspot.

Several additional comments about the spot term in Eqn.~(\ref{basicModel}) are in order. First, this term assumes that the spot is visible around the entire orbit. Roughly speaking, this is only true for spots that are located at places on the star such that $i + \alpha < 90^\circ$, where $i$ is the orbital inclination and $\alpha$ is the colatitude (defined as the complement of the latitude) of the spot, assuming that the spot is located on the hemisphere centered on the $\alpha=0$ pole (Tran et al.~2013). As was argued in Tran et al.~(2013), empirically it appears that this criterion is largely met in those systems where the $O-C$ curves are clearly anticorrelated. Second, the zero of longitude is defined so that it contains the line joining the two stellar centers (near the inner Lagrange point). Third, the sign of the spot term assumes that spots are on the farther star at lightcurve minimum.  If the spot is on the nearer star, the sign of the coefficient must be interpreted accordingly. Finally, if the starspot moves in longitude uniformly with time, due to differential stellar rotation or due to imperfectly synchronized stellar rotations, then $\ell$ will vary linearly with time.  Wandering of the spot on the stellar surface may produce non-periodic variations in time.

{ We also note that a bimodal starspot distribution in longitude on one of the stars could introduce a signal at twice the orbital frequency, potentially contributing to, and interfering with, the $\cos 2 \omega t$ term in Eqn.~(\ref{basicModel}).  While the Fourier harmonics of starspots do tend fall quickly with frequency (see, e.g., Rappaport et al.~2014), this is a potentially important effect to watch for.  Empirically, however, in this work we do not typically find an important time-varying component to the $\cos 2 \omega t$ term.}

\begin{figure}
\begin{center}
\includegraphics[width=\columnwidth]{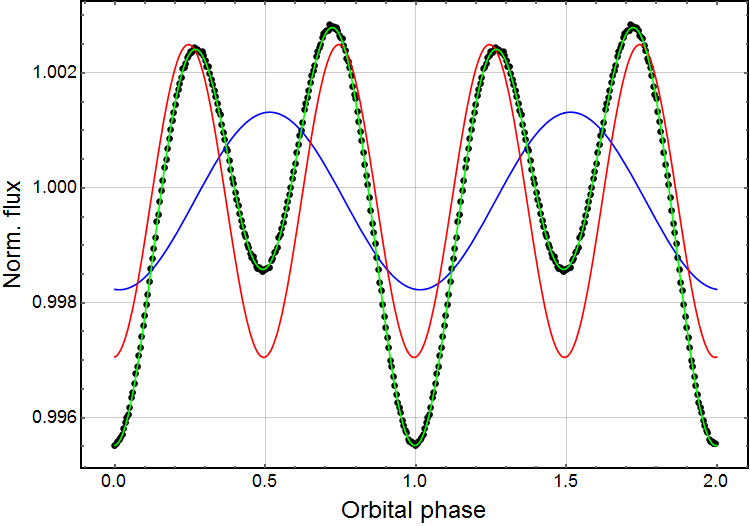}
\caption{Folded lightcurve for KIC 9002076 (black points) with a fit to a function given by Eqn.~(\ref{basicModel}) with $S$ set equal to zero (i.e.~no average contribution from star spots). The blue, red, and green curves show the $\omega t$ and $2\omega t$ terms as well as their sum, respectively, where $\omega$ is the orbital frequency. Phase zero is taken to be at the primary minimum in the folded light curve. Evidently, when spot behavior is taken out of the picture, the remaining terms---one sinusoid at the binary frequency $\omega$, one at $2\omega$, and a constant---in Eqn.~(\ref{basicModel}) are sufficient to describe the lightcurve quite accurately.}
\label{foldPlot}
\end{center}
\end{figure}

We demonstrate how well the first three terms of the expression in Eqn.~(\ref{basicModel}) can characterize the basic orbital light curve by fitting them to the folded light curve for one of the short-period binaries in this study, KIC 9002076 (see Fig.~\ref{foldPlot}). The process of folding the data tends to average out the starspot behavior, because $\ell$ varies as the spot migrates (as we show in this work) and thus generally takes on a wide range of values over a long time interval. Figure \ref{foldPlot} displays the folded light curve for KIC 9002076 which has (i) very low amplitude modulations, and (ii) two different light curve minima whose depths differ by roughly a factor of two. The fitted curves show the combined contributions of the first three terms in Eqn.~(\ref{basicModel}). As can be seen, the sum of the three terms fits the overall folded lightcurve extremely well.

\subsection{The Spot Term}
\label{spotTerm}

\begin{figure}
\begin{center}
\includegraphics[width=\columnwidth]{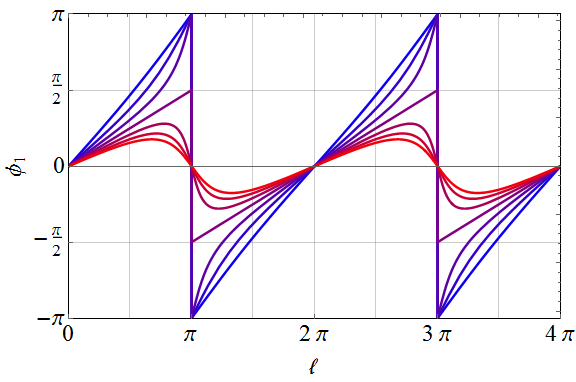}
\vglue0.5cm
\includegraphics[width=\columnwidth]{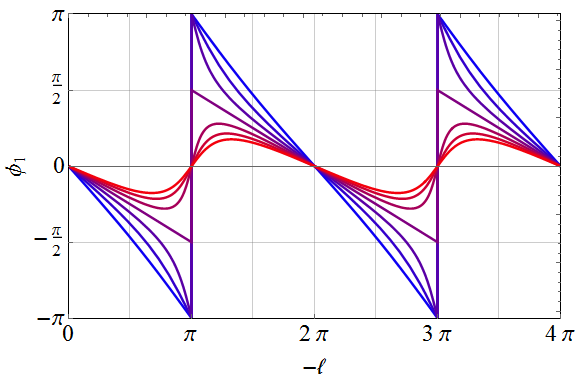}
\caption{A plot of $\phi_1$ as given by Eqn.~(\ref{fitconst}) for several values of $S/E_1$, from 0.5 (red) to 10 (blue). This ratio is effectively the ratio of the spot amplitude to the difference between primary and secondary minima depths. The upper plot shows $\phi_1$ as $\ell$ increases with time, while the lower plot shows the shape of $\phi_1$ if $\ell$ is instead a decreasing function of time. Notice that if $S/E_1$ is large (bluer curve), then $\phi_1$ behaves more linearly, with phase wrapping; if $S/E_1$ is small, $\phi_1$ varies almost sinusoidally, with much smaller (and better defined) amplitude. Nonetheless, whether the phase curve exhibits a slow rise followed by a steeper decline, or vice versa, indicates whether the spot motion is prograde or retrograde.}
\label{arctan}
\end{center}
\end{figure}

We next examine how well the expression on the right in Eqn.~(\ref{basicModel}), including and especially the $S \cos{(\omega t + \ell)}$ term, can fit the flux data from a contact binary with migrating spots. If we assume that the value of $\ell$ in Eqn.~(\ref{basicModel}) does not change rapidly, we may take it to be a constant for short time intervals. In that case we can add the two terms at frequency $\omega$ in Eqn.~(\ref{basicModel}) and write it as:
\begin{equation}
f(t) \simeq C - a_1 \cos{(\omega t + \phi_1)} - a_2 \cos{(2\omega t+\phi_2})
\label{fitFunction}
\end{equation}
where, as above, $t$ is the time relative to the time of a primary minimum, and therefore $\phi_2$, which we have included for robustness of the fit, should remain near zero.
While $a_2$ is simply equal to $E_2$, the amplitude $a_1$ and phase $\phi_1$ are related to $E_1$, $S$, and $\ell$ by
\begin{align}
a_1 & = \sqrt{S^2+E_1^2+2SE_1 \cos{\ell}} \\
\tan{\phi_1}& = \frac{S\sin{\ell}}{E_1 + S\cos{\ell}}
\label{fitconst}
\end{align}
In the limit that the starspot amplitude dominates the orbital amplitude at the orbital frequency, i.e.~$S \gg E_1$, then $\phi_1 \simeq \ell$. In that case, as the spot moves all the way around the star, $\phi_1$ will cycle through a net angle of $2 \pi$ radians. If instead $E_1 \gg S$, then $\tan \phi_1 \simeq (S/E_1)\sin \ell$, and the value of $\phi_1$ will oscillate over an interval much smaller than $2\pi$ radians even as $\ell$ changes by $2 \pi$ radians. These behaviors are summarized in Figure \ref{arctan}, where we plot $\phi_1$ as given in Eqn.~(\ref{fitconst}) for various ratios of $S/E_1$.

The results of Tran et al.~(2013) showed that the spot longitude, $\ell$, varies significantly even if slowly given the long duration of almost all of the {\em Kepler} light curves. Therefore we fit the function in Eqn.~(\ref{fitFunction}) to successive segments of the data, arbitrarily chosen to be 500 {\em Kepler} LC samples in duration ($\sim$10 days, or between $\sim$10 and $\sim$30 binary orbital periods).  The fits are carried out for segments with starting points offset by 100 LC intervals ($\sim$2 days) with respect to one another. The time variation of the fitted parameters can be used to infer changes in spot longitudes.

\begin{figure*}
\begin{center}
\includegraphics[width=.9\textwidth]{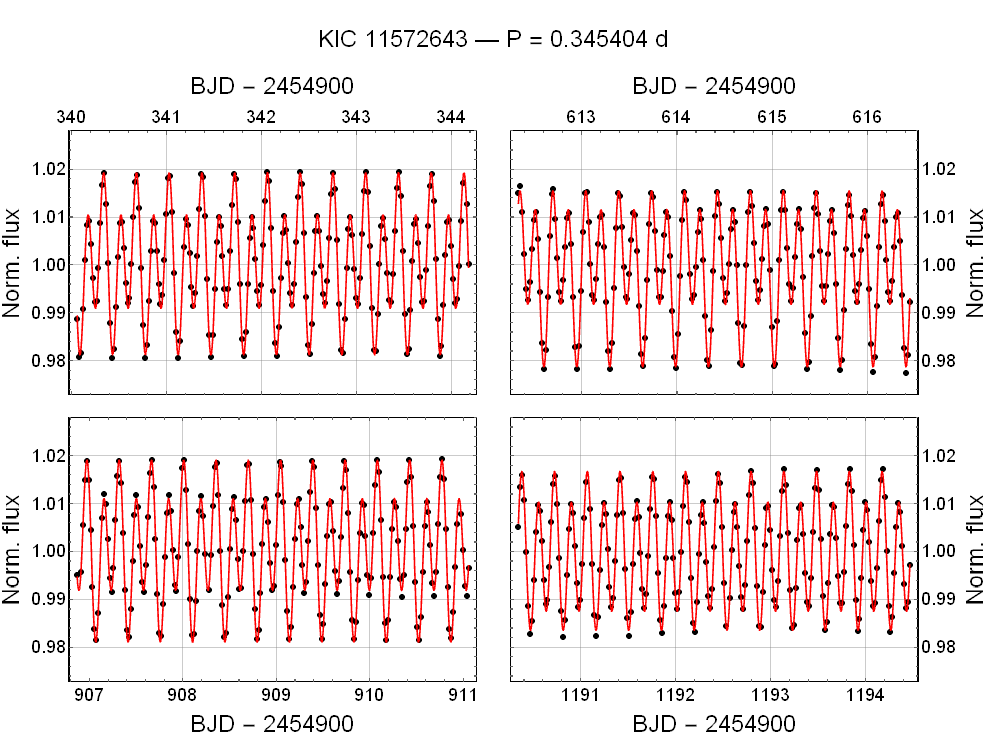}
\caption{An illustrative example of how well the fitting function given by Eqn.~(\ref{fitFunction}) can describe the {\em Kepler} data for a contact binary with an assumed migrating starspot.  Here we show the results for four different segments of the flux data from the binary KIC 11572643. In {\em each} segment a single value for the spot amplitude and phase is assumed; these parameters do vary from segment-to-segment.  The shape of the fitting function (in red) is noticeably different in each segment, presumably due to the changes in spot longitude, but the fit is still excellent in all four segments of data.}
\label{severalFits}
\end{center}
\end{figure*}

The ability of the function in Eqn.~(\ref{fitFunction}) to fit 500-LC sample intervals of a light curve that changes quite obviously over longer timescales is demonstrated by results for one of the contact binaries in our study, KIC 11572643. Fig.~\ref{severalFits} shows four short segments of the data each separated by $\sim$300 days together with the best-fit functions. Even though the relative depths of the minima are clearly different in each segment, the simple model fits rather well and enables the spot phase and amplitude to be extracted.

\subsection{Diagnostic Dissection of a Light Curve}
\label{params}

\begin{figure*}
\begin{center}
\includegraphics[width=0.80\textwidth]{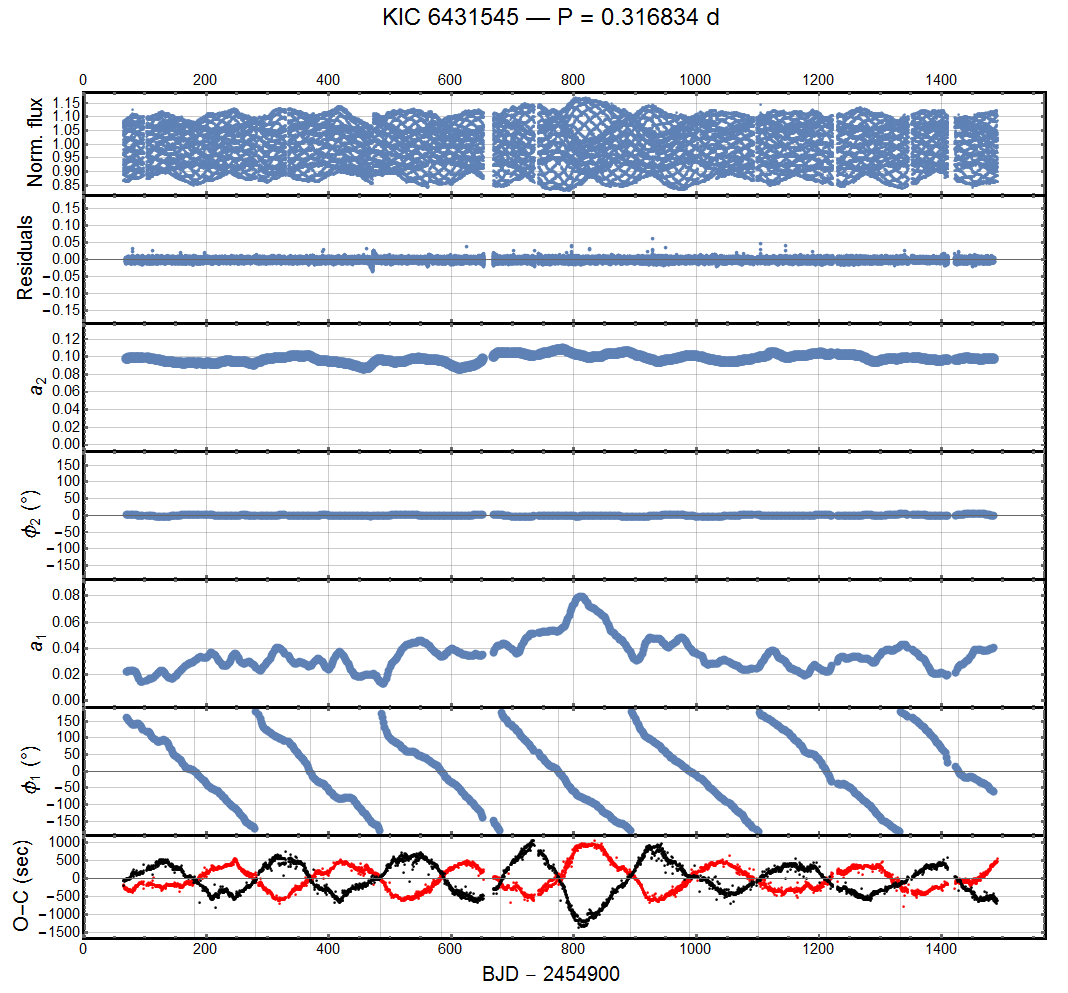}
\caption{Parameters determined by fits of a function of the form given by Eqn.~(\ref{fitFunction}) to 500-point segments (which are then stepped by 100 points) of the light curve of the contact binary KIC 6431545. The panels are as follows (top to bottom): the normalized {\em Kepler} flux data; the residuals after the fits have been subtracted from the flux data; individual fitted parameters $a_2$, $\phi_2$, $a_1$, and $\phi_1$; and the $O-C$ curve for the primary minimum (red points) and secondary minima (black points). The vertical gridlines in the bottom two panels are placed at the locations where $\phi_1$ crosses 0 and $\pi$, showing that these locations match with those where the $O-C$ curves intersect each other.  Note that the flux data (top panel) exhibit a Moir\'e pattern, i.e., a beat between the binary period and the {\em Kepler} long-cadence sampling time.}
\label{diagnosticFigure}
\end{center}
\end{figure*}

Figure \ref{diagnosticFigure} illustrates the decomposition of an entire {\em Kepler} light curve for one of the short-period binaries in our study, KIC 6431545, into various amplitudes and phases as a function of time.

The top panel provides an overview of the {\em Kepler} lightcurve, including a sense of the overall variation of the peak-to-peak amplitude. The second panel shows the residuals after the best fit for each segment has been subtracted; since we step the data segments that we fit by 100 points each, the residuals for each consecutive series of 100 points are based on a slightly different fit. The results in this panel give a direct sense of the overall performance of the fitting function since the vertical scale has the same range as that of the light curve plot. The next four panels show the best-fit values of the parameters $a_2$, $\phi_2$, $a_1$, and $\phi_1$ (defined as in Eqn.~(\ref{fitFunction})).

The best-fit values of $a_2$ and $\phi_2$ (third and fourth panels) vary very little, indicating that the corresponding $2\omega t$ term remains relatively consistent throughout the light curve. Since this is the dominant term in the description of the orbital contribution to the light curve, this behavior is as expected, assuming (i) there are no external perturbations to the orbit such as those that would be produced by a third body in the system, { or (ii) a dominant bimodal starspot distribution in longitude causing unanticipated power in the $\cos 2 \omega t$ term.}  The fitted values of $a_1$ (fifth panel) indicate the amplitudes of the flux modulations caused by the starspot (over and above the nominal binary lightcurve). Of note in this particular system is that $a_1$ is largest where the amplitudes of the flux data themselves and the $O-C$ curves (bottom panel) are both also largest. This expected correlation is indeed visible in many of the sources we analyzed.

The sixth panel in Fig.~\ref{diagnosticFigure} is striking. It shows that $\phi_1$ decreases at a nearly steady rate, wrapping around $2\pi$ radians presumably in concert with the spot longitude. This behavior indicates that the actual frequency of the spot term is slightly {\em lower} than the orbital frequency. This would be expected from the variations due to a spot on a star that is rotating at a rate slightly {\em below} that of the binary orbit. According to this interpretation, the $\sim$6.5 cycles of $\phi_1$ indicate that the spot circles the star (in a frame of reference rotating with the binary) just over 6.5 times during the four years of the {\em Kepler} observations.

It is important to note that the slow drift of $\phi_1$ almost certainly arises from the spot being carried around the star as viewed in a reference frame rotating at the orbital frequency. This would occur if the entire star is rotating like a solid body at a rate below or above that of the orbit, i.e., if the rotation of the star with the spot is not synchronized with the orbital motion (i.e., not tidally locked).  A second possibility is that the rotation of part of the star, likely the equatorial region, is indeed tidally locked to the orbit, but that there is nevertheless differential rotation so that there is a region at, e.g., a higher stellar latitude that is not rotating at the orbital rate.  { A third possibility is that the spot migrates in longitude due to another cause, one possibility being a systematically time-varying magnetic field configuration that might drive the location where spots appear.}  Of course, some combination of these is also possible. In { any case, the observed} nonuniformities in the drift rate of the spot are likely to be due to motions of the spot relative to the mean motion of the stellar surface at nearby latitudes. 

The bottom panel in Fig.~\ref{diagnosticFigure} shows the $O-C$ curves for the primary (red points) and secondary (black) lightcurve minima.

\begin{figure*}
\begin{center}
\includegraphics[width=0.9\textwidth]{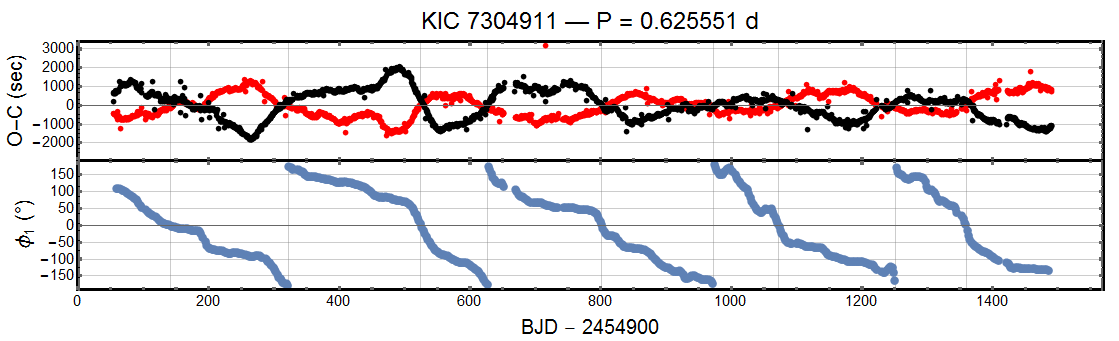}
\vglue0.5cm
\includegraphics[width=0.9\textwidth]{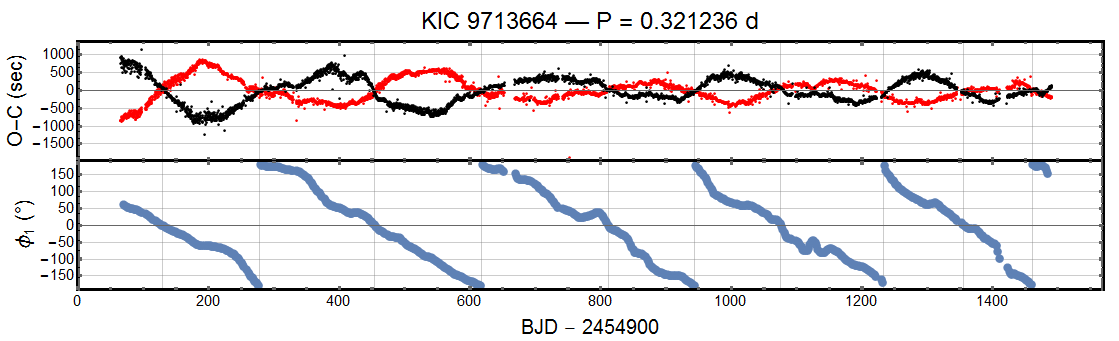}
\vglue0.5cm
\includegraphics[width=0.9\textwidth]{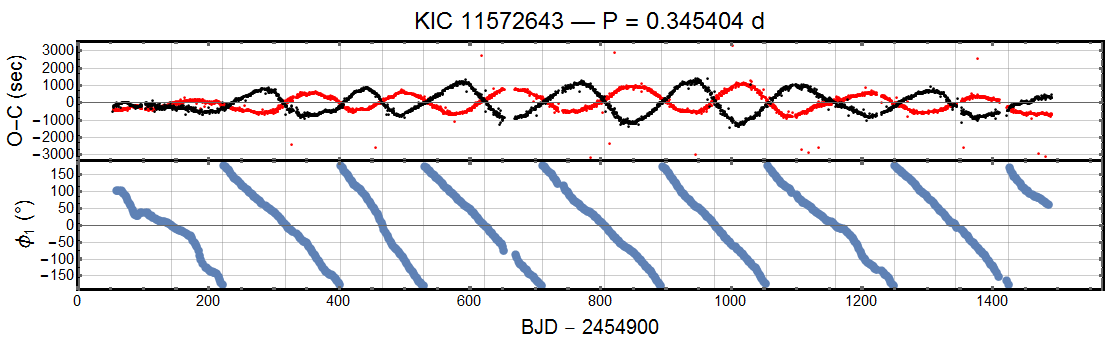}
\caption{The first 3 of 12 examples of binaries for which we compare the $O-C$ and $\phi_1$ curves (KIC 7304911, KIC 9713664, and KIC 11572643). For each binary, the upper plot shows the $O-C$ curves for the primary (red) and secondary (black) minima. The lower plot is of $\phi_1$, as defined in Eqn.~(\ref{fitFunction}); we believe $\phi_1$ tracks the location of a starspot (or group of starspots) on one of the binary stars.  The KIC number and the orbital period are given at the top of the plots for each system. The vertical gridlines are placed where the plot of $\phi_1$ intersects 0 or $\pi$. (The zero-point reference for $\phi_1$ is taken to be the mean value of $\phi_2/2$, when averaged over all the {\em Kepler} data.) Note how well the $O-C$ intersections agree with these gridlines -- strongly indicating that both types of curves are tracking the effect of the same phenomenon in the lightcurves, namely spots.}
\label{fig:phiOC1}
\end{center}
\end{figure*}

\section{Phasing the $O-C$ With the $\phi_1$ Curves}
\label{phasingUp}

Starspot behavior is manifest in both $O-C$ curves and plots of $\phi_1$ versus time (as defined above), although there are significant differences.  To be specific, the $O-C$ values measure timewise-local (near primary or secondary minimum) spot photometric effects, whereas phase tracking involves fits of entire orbital cycles. Thus, it is natural to check how the two are related.

One way to ascertain whether the $O-C$ and $\phi_1$ curves are measuring effects due to the same starspots is to check how the zeros of the $\phi_1$ plots (i.e., $\phi_1 = 0$) align with the intersections of the primary and secondary $O-C$ curves.\footnote{The intersections of the primary and secondary $O-C$ curves would occur at zero delay as well, except for any \emph{correlated} behavior between the primary and secondary curves due to extraneous causes.} However, we must first be sure that the meaning of ``phase'' is identical between the two plots. In the $O-C$ curves, the points of zero deviation for both the primary and secondary minima occur (by symmetry) when the spot is at a stellar longitude $\ell = 0$ or $\pi$, which according to Eqn.~(\ref{fitconst}) implies $\phi_1 = \ell$ at that point. Thus, the zeros in the $O-C$ curves should align with values of $\phi_1$ equal to 0 or $\pi$.

In connection with Eqn.~(\ref{basicModel}), it was noted that $t=0$ should be set to the time of a primary minimum.  If the times of the primary minima are not precisely periodic, there is latitude in the choice of the time origin. Indeed, the choice of the time origin was acknowledged to be approximate in Eqn.~(\ref{fitFunction}) by allowing a nonzero value of $\phi_2$.  In practice, the overall average value of $\phi_2$ was used as a reference to effectively set the time of orbital phase zero, and to thereby also set the zero-point reference for $\phi_1$, i.e., the figures show the values of $\phi_1 - \langle \phi_2 \rangle / 2$ and $\phi_2 - \langle \phi_2 \rangle $ rather than the raw values of $\phi_1$ and $\phi_2$.  

\null{} \vspace{0.1cm}

\subsection{Comparison of $O-C$ and $\phi_1$ Curves for a Sample of a Dozen Binaries}

From a sample of 414 short-period binaries with anticorrelated $O-C$ curves listed in Table \ref{tab:Period} and Appendix A\footnote{Tran et al.~(2013) first identified 390 of this set of short-period binaries with anticorrelated $O-C$ curves.}, we selected 12 that had especially well-behaved $O-C$ curves. In Sect.~\ref{collresults} we will present a summary of the $O-C$ periodicities for the entire sample of systems. Here the $O-C$ and $\phi_1$ curves for the 12 selected binaries are compared in detail.

\begin{figure*}
\begin{center}
\includegraphics[width=0.9\textwidth]{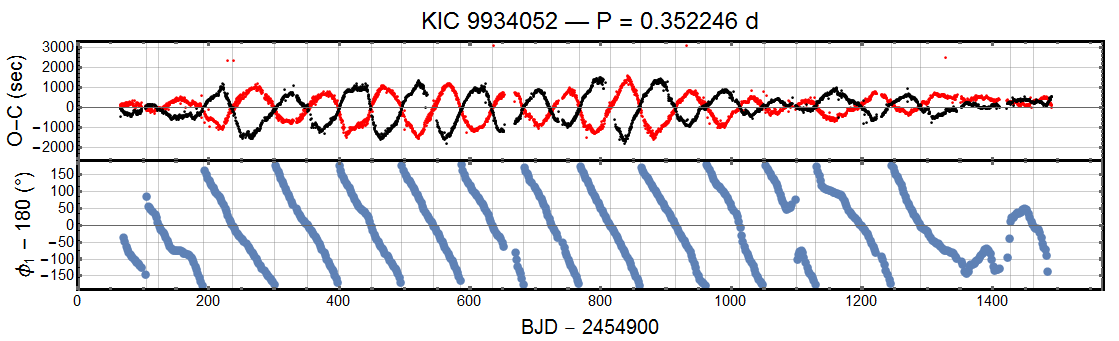}
\vglue0.5cm
\includegraphics[width=0.9\textwidth]{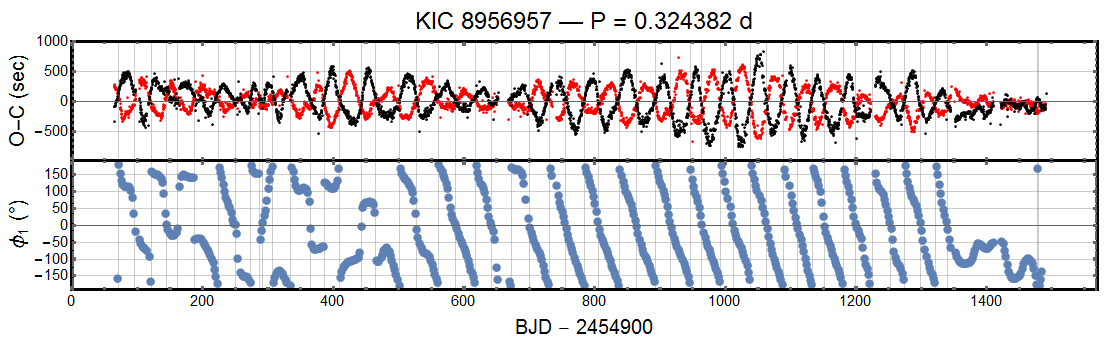}
\vglue0.5cm
\includegraphics[width=0.9\textwidth]{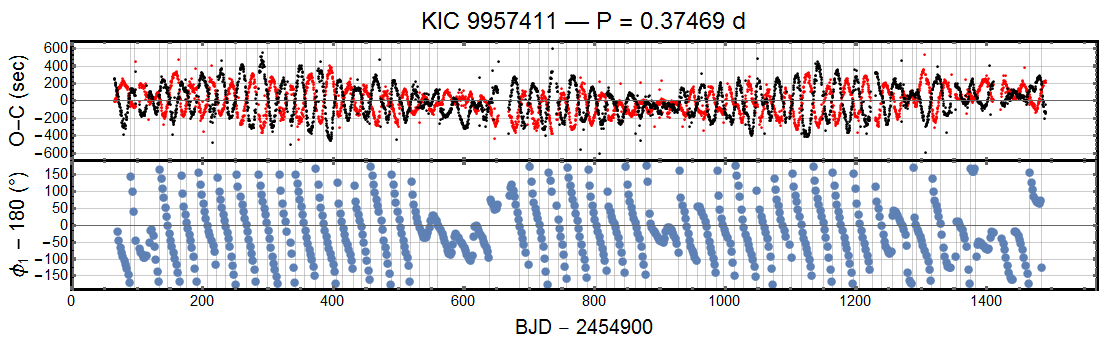}
\caption{The second group of 3 of 12 examples of binaries displayed similarly to Fig.~\ref{fig:phiOC1}. The binaries chosen for this figure exhibit more rapid rates of change in $\phi_1$ than those shown in Fig.~\ref{fig:phiOC1}, which may be taken to indicate higher speeds of spot motion in longitude ($\phi_1$ is the spot longitude relative to a line joining the stellar centers). The $\phi_1$ curves for KIC 9934052, KIC 8956957, and KIC 9957411 wrap around by an angle of 2$\pi$ radians approximately 13, 21, and 31 times during the {\em Kepler} observations, corresponding to spots migrating all the way around their host star numerous times. There are several intervals in each of these $\phi_1$ curves where the phase ceases to wrap around in a regular fashion---these correspond to intervals of weak (low-amplitude) spot behavior and correspondingly low amplitudes in the $O-C$ curves.  The $\phi_1$ curves for the top and bottom panels are presented as ``$\phi_1 - 180$'' in order to keep visible any clearly oscillatory or pseudo-oscillatory behavior. We do likewise in the rest of this paper for any other binaries for which this presents a clearer picture.}
\label{fig:phiOC2}
\end{center}
\end{figure*}

\begin{figure*}
\begin{center}
\includegraphics[width=0.9\textwidth]{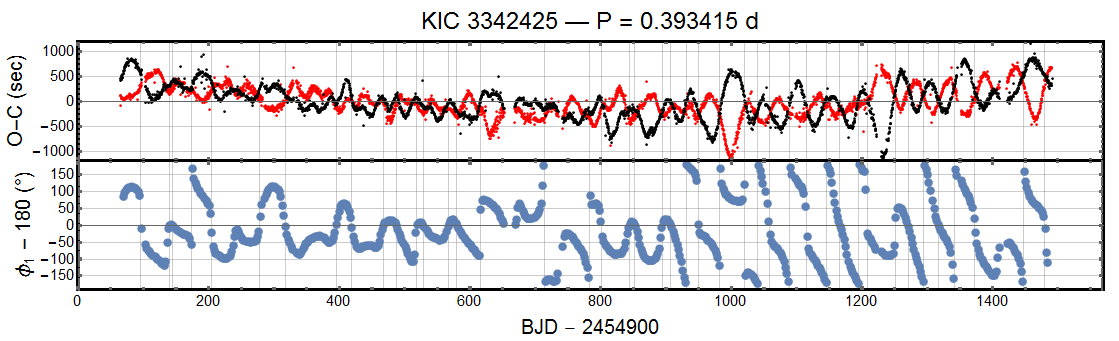}
\vglue0.5cm
\includegraphics[width=0.9\textwidth]{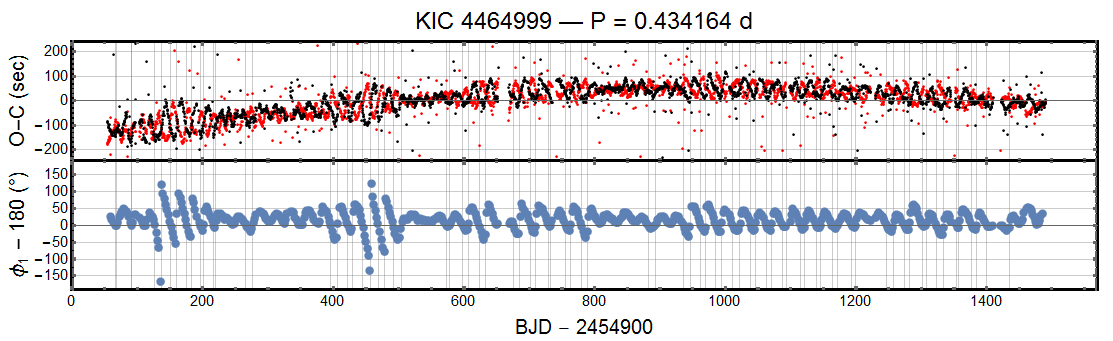}
\vglue0.5cm
\includegraphics[width=0.9\textwidth]{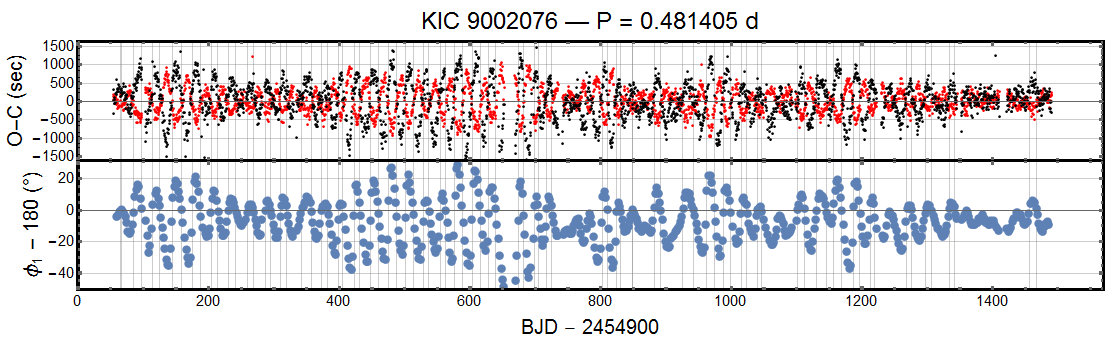}
\caption{The third group of 3 of 12 examples of binaries displayed similarly to Fig.~\ref{fig:phiOC1}. These three binaries (KIC 3342425, KIC 4464999, and KIC 9002076) demonstrate the effect of having a spot amplitude that is comparable to, or smaller than, the difference between the depths of the primary and secondary minima (see Eqn.~(\ref{fitconst}) and the accompanying discussion in Sect.~\ref{spotTerm}). In these cases, $\phi_1$ oscillates around $\pi$ (we present $\phi_1 - \pi$ to make the oscillations clear), rather than behaving linearly modulo $2\pi$. Of course, $\phi_1$ could oscillate around 0 instead---the offset by $\pi$ is only a question of which star the spot is on. Notice that the phase curves all consistently have steep rises followed by slower declines, indicative of spots that rotate more slowly than the orbit. Also note that the vertical scale on the $\phi_1$ plot for KIC 9002076 is expanded to better show the oscillations.}
\label{fig:phiOC3}
\end{center}
\end{figure*}

\begin{figure*}
\begin{center}
\includegraphics[width=0.9\textwidth]{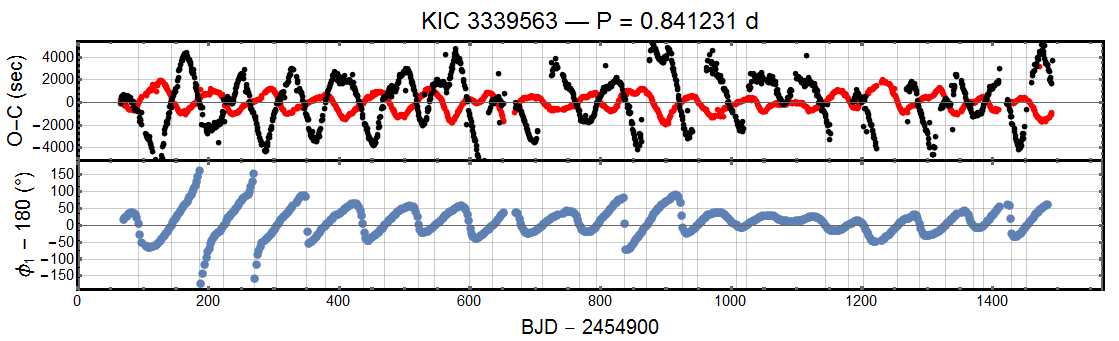}
\vglue0.5cm
\includegraphics[width=0.9\textwidth]{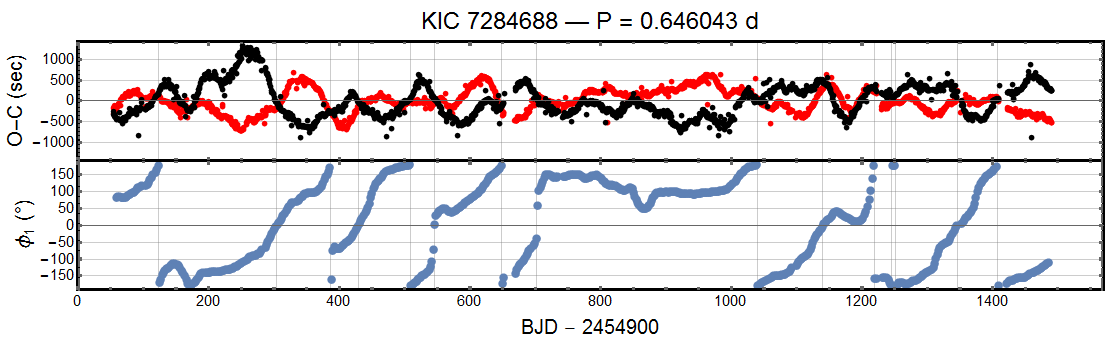}
\vglue0.5cm
\includegraphics[width=0.9\textwidth]{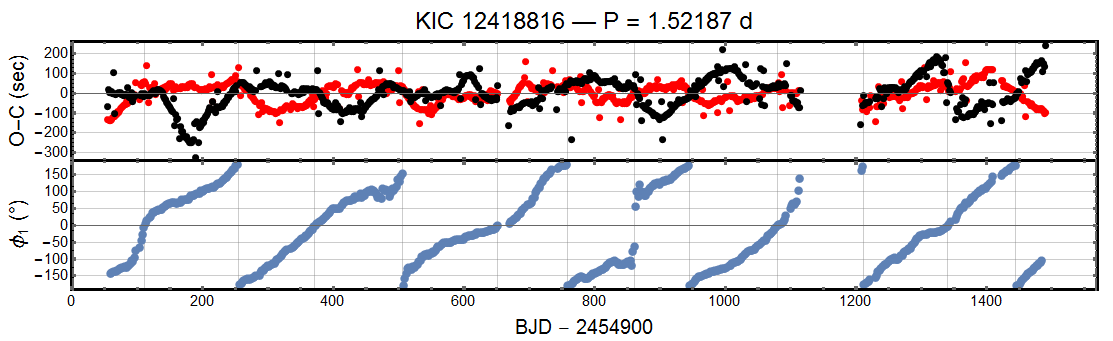}
\caption{The fourth group of 3 of 12 examples of binaries displayed similarly to Fig.~\ref{fig:phiOC1}. The three binaries shown here (KIC 3339563, KIC 7284688, and KIC 12418816) are all examples for which $\phi_1$ is predominantly increasing (aside from wrapping around by $2\pi$) with time, i.e., the differential rotation of the spots is prograde with respect to the orbit.  In the case of KIC 3339563 this behavior is clear because of the slow rises and steep falls during the oscillatory portion of the $\phi_1$ curve which dominates most of the data train, as well as in the shorter wrapping (non-oscillatory) interval near the start. In all other respects, these curves are similar to the other 9 sources as shown in Figs.~\ref{fig:phiOC1}, \ref{fig:phiOC2}, and \ref{fig:phiOC3}.}
\label{prograde}
\end{center}
\end{figure*}

Figure \ref{fig:phiOC1} shows the $O-C$ and $\phi_1$ curves for  { KIC 7304911, KIC 9713664, and KIC 11572643.}  In all three cases, the pair of curves shows excellent consistency in the sense that the times when the primary and secondary $O-C$ curves  intersect coincide with the times when $\phi_1 = 0$ or $\pi$. This is quite compatible with our interpretation that both the $O-C$ behavior and $\phi_1$'s cycling through 2$\pi$ are due to spots drifting slowly in longitude { (with respect to a line joining the two stellar centers).} Note also that in all three of these cases, $\phi_1$ is basically linear modulo $2\pi$---this is because these sources have very similar depths for the primary and secondary lightcurve minima.  As a result, $E_1$ is negligible, and per the discussion in Sect.~\ref{spotTerm}, $\phi_1 = \ell$, meaning that it directly tracks the spot longitude.

In all three cases shown here, as is true for some 34\% of the 414 systems we investigated (see Table \ref{tab:Period} and Appendix A), the phases decrease essentially monotonically with time (modulo $2\pi$), which is indicative of a slowly counter-rotating spot (i.e., one with retrograde motion). About 13\% of the systems instead exhibited $\phi_1$ {\em increasing} with time (i.e., prograde motion), while the remaining 53\% displayed either erratically moving or relatively stationary phases (the latter being those that did not wrap by $2\pi$ more than once over the entire duration of the light curve). The percentages of systems with different spot rotational behaviors are summarized in Table \ref{percents}.

\begin{deluxetable}{lc}
\tablecolumns{2}
\tablewidth{0pt}
\tablecaption{Spot Movement Behavior Among 414 Contact Binaries\label{wrappingDistr}}
\tablehead{Classification & Percentage}
\startdata
Prograde Motion & 13.0\\
Retrograde Motion & 34.3\\
Small Phase Changes & 45.6\\
Erratic Behavior & 7.2
\enddata
\label{percents}
\end{deluxetable}

The $O-C$ and $\phi_1$ curves for KIC 9934052, KIC 8956957, and KIC 9957411 are displayed in Fig.~\ref{fig:phiOC2}. { These three systems clearly exhibit spot motions that are substantially more rapid than those of the examples shown in Fig.~\ref{fig:phiOC1}. The $\phi_1$ curves wrap around by an angle of 2$\pi$ radians substantially more times in 1400 days than for any of the systems shown in Fig.~\ref{fig:phiOC1}. All three of these $\phi_1$ curves} contain intervals, aligned with intervals of low-amplitude $O-C$ activity, in which $\phi_1$ fails to wrap around by 2$\pi$ and instead oscillates. Both of these behaviors, i.e., regions of low-amplitude $O-C$ curves and failure of $\phi_1$ to wrap, indicate weak starspots, the former from our discussion in Sect.~\ref{spotTerm} and the latter from Tran et al.~(2013).

Oscillatory (non-wrapping) $\phi_1$ behavior is seen in the results for KIC 3342425, KIC 4464999, and KIC 9002076 (see Fig.~\ref{fig:phiOC3}).  From this, we conclude, again based on the discussion in Sect.~\ref{spotTerm}, that the spot amplitude $S$ is comparable to, or smaller than, the amplitude $E_1$ (see Eqn.~(\ref{fitconst})), and therefore the variation of $\phi_1$ is limited to a range of only $\sim\pm S/E_1$ radians. Indeed, among the systems we have studied, those where the depths of the two light-curve minima are significantly different and $E_1$ is non-negligible tend to have low-amplitude oscillating $\phi_1$ curves. In spite of their small amplitudes, the phase $\phi_1$ of the spots can be tracked rather well (with zeros still matching up quite well with intersections in the primary and secondary $O-C$ curves), and we see clearly that the phase curves consistently have sharper rises followed by slower falls. This provides an indication, in the absence of wrapping, of spots that rotate more slowly than the orbit (see Fig.~\ref{arctan}).

We present the final three examples in Fig.~\ref{prograde} primarily to show that there do exist binaries with spots that are apparently rotating {\em faster} than the orbit; otherwise they share characteristics that are exhibited by the other 9 examples.  The overall ratio of clearly prograde systems, such as those in Fig.~\ref{prograde}, to clearly retrograde systems is $\sim$4:10.  

\begin{figure*}
\begin{center}
\includegraphics[width=.9\textwidth]{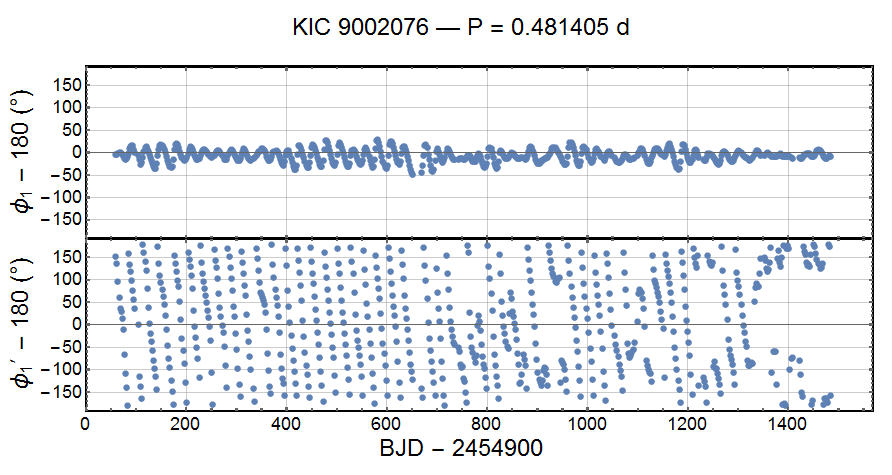}
\caption{Plots of $\phi_1$ before (above) and after (below) subtracting an overall $\langle a_1\rangle \cos{\omega t}$ term from the lightcurve for KIC 9002076. (To mark its distinctness from $\phi_1$, we label the quantity in the lower figure $\phi_1^\prime$.) By fitting Eqn.~(\ref{fitFunction}) to the whole data series at once, the spot behavior averages out, as it does in the folded light curve shown in Fig.~\ref{foldPlot}. Then, subtracting the fitted $\langle a_1\rangle \cos{\omega t}$ term is equivalent to subtracting the $E_1 \cos{\omega t}$ term in Eqn.~(\ref{basicModel}). There are no significant differences between the depths of the primary and secondary minima in the resulting light curve, so spot behavior is more clearly picked up by new, segment-by-segment fits using Eqn.~(\ref{fitFunction}).}
\label{compareSubtraction}
\end{center}
\end{figure*}

To verify that a system where $\phi_1$ varies over only a small range of angles (see, e.g., Fig.~\ref{fig:phiOC3}) is, in fact, consistent with our model, the analysis must be redone after the $E_1 \cos \omega t$ variation has been removed from the light curve. This is straightforward to accomplish. We simply subtract off an approximation to the $E_1 \cos \omega t$ term from the data, having taken $E_1$ to be the value of $a_1$ from the fit of Eqn.~(\ref{fitFunction}) to the {\em entire data set}.  This is equivalent to fitting the folded lightcurve, as in Fig.~\ref{foldPlot}, and, in the process, this tends to average out the spot contribution as noted above.  Thus subtracting this fitted term removes the difference between the depths of the primary and secondary minima. We apply this method to KIC 9002076; the results are shown in Fig.~\ref{compareSubtraction}. The top panel is a copy of the bottom panel in Fig.~\ref{fig:phiOC3}, where no $E_1 \cos \omega t$ term was subtracted before the phase fitting was done. The bottom panel shows the corresponding plot of $\phi_1$ vs.~time made after the $E_1 \cos \omega t$ term was subtracted from the light curve. Note that after the subtraction, $\phi_1$ wraps around 2$\pi$ radians many times.

\section{Summary Results for 414 Binaries with Anticorrelated $O-C$ Curves}
\label{collresults}

The $O-C$ curves contain information about the timescales of the motions of the spots around the stars in the rotating frames of the binaries.  For the estimation of the timescales, it is best to use the difference of the primary and secondary $O-C$ curves for each system so as to magnify the anticorrelated effects and eliminate correlated ones. Simple interpolation was used to obtain the difference curves.  Then, the autocorrelation function, $ACF(t)$, was computed for each $O-C$ difference series in order to extract the underlying periodicities (see, e.g., McQuillan et al.~2013). To accomplish this, we fit the autocorrelations over delays of 0 to 400 days with the somewhat arbitrary analytic function
\begin{equation}
ACF(t) \simeq \frac{A\cos{Bt}}{(t+1\,{\rm d})^C}
\label{autocorrFit}
\end{equation}
with constants $A$, $B$, and $C$. The characteristic starspot migration period is then taken to be $P_{\rm mig} = 2\pi/B$.\footnote{The spot migration period is approximately the modulation period in the $O-C$ curve. See Eqn.~(7) in Tran et al.~(2013)} This is the time it takes a spot to move completely around the host star in the reference frame of the rotating binary. If $P_s$ is defined as the spot rotation period as viewed from inertial space, then these two periods are related to the binary orbital period, $P_{\rm orb}$ by:
\begin{equation}
P_{\rm mig} = \frac{P_{\rm orb} P_s}{| P_{\rm orb}-P_s |} \simeq \frac{P_{\rm orb}}{| 1-P_s/P_{\rm orb} |}
\label{TranEq}
\end{equation}
as long as $P_s$ and $P_{\rm orb}$ are close in numerical value. The absolute value must be taken when $P_s > P_{\rm orb}$ in order to keep $P_{\rm mig}$ positive.  We further define the quantity $\kappa$ by
\begin{equation}
\kappa \equiv \frac{P_{\rm s}}{P_{\rm orb}}-1
\label{kappa}
\end{equation}
and its absolute value is then given by
\begin{equation}
|\kappa| = \frac{P_{\rm orb}}{P_{\rm mig}} .
\label{abskappa}
\end{equation}
Based on sunspot behavior, a simple model for the differential rotation of spotted stars has been adopted for this problem (e.g. Hall \& Busby 1990; Kalimeris 2002) such that {
$$\kappa' \equiv \frac{P-P_{\rm eq}}{P_{\rm eq}}= k \cos^2 \alpha$$
where $P$ and $P_{\rm eq}$ are the rotation periods of the star at colatitude $\alpha$ (as defined above)} and at the equator, respectively.  The differential rotation constant, $k$, in our application, may be positive or negative or zero. In the Sun, $k$ is positive and and $\kappa$ describes the increasing nature of the spot rotation period with increasing latitude (decreasing colatitude). In our close binary problem, the details of any differential rotation and the exact cause of the apparent spot motion are not known, so we simply use $\kappa$ as the parameter to be measured; $k$ and $\kappa$ may have either sign. In relating $k$ to $\kappa$, we note that 
the quantity $\cos \alpha$ is likely to be $\gtrsim 0.7$ or $\cos^2 \alpha \gtrsim 0.5$. This results from the various constraints put forth by Tran et al.~(2013) for the orbital inclination and spot colatitude in order to accommodate the observed anticorrelated $O-C$ curves. Stated simply, this results from the requirements that $i \gtrsim 40^\circ$ and $i + \alpha < 90^\circ$, and therefore $\alpha \lesssim 50^\circ$.

\begin{figure}
\begin{center}
\includegraphics[width=\columnwidth]{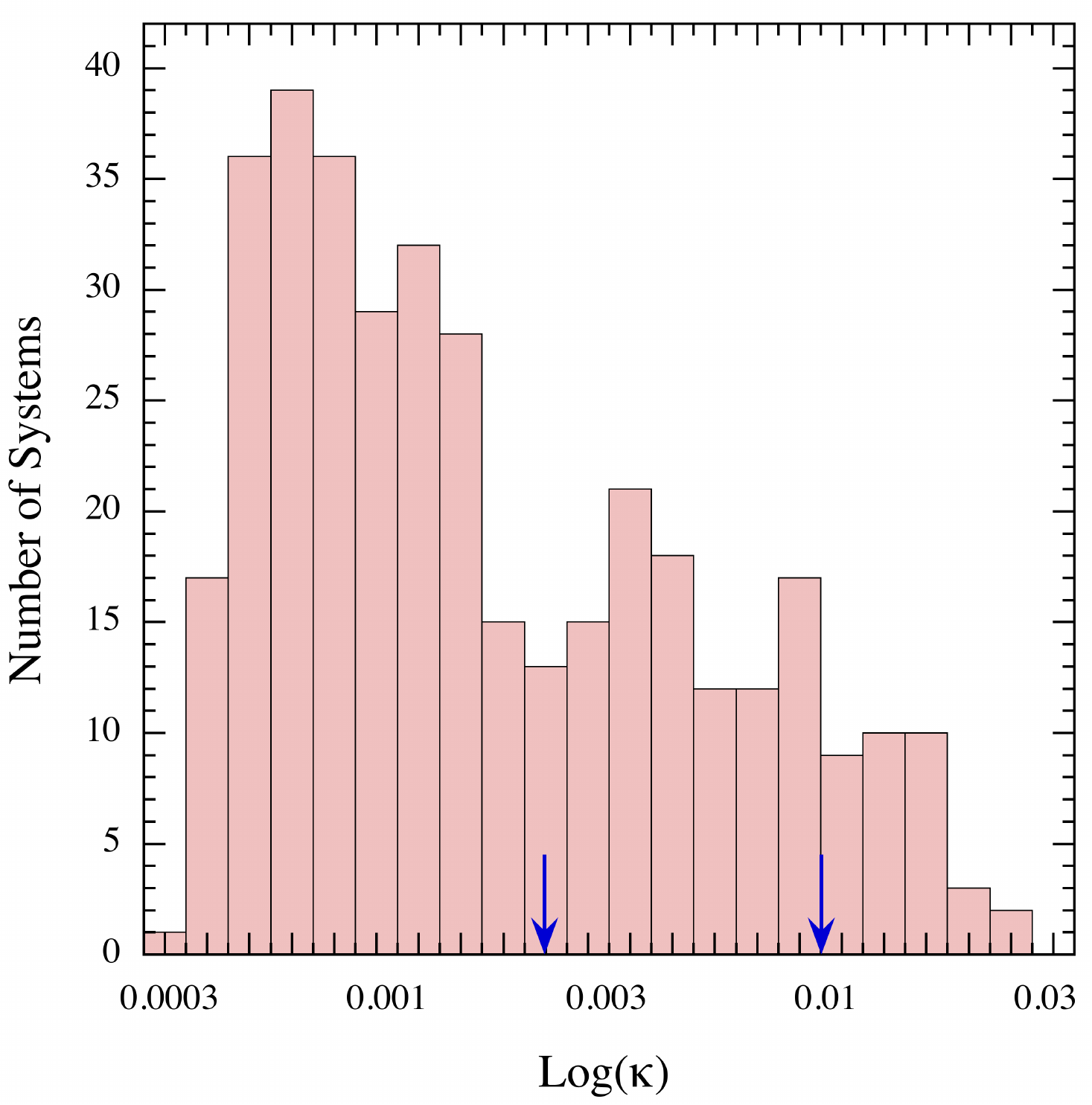}
\caption{A histogram of our calculated differential rotation coefficient $|\kappa|$ (as defined in Eqn.~\ref{abskappa}) for each of 390 contact binaries with anticorrelated primary and secondary $O-C$ curves.  The bins in $|\kappa|$ are spaced logarithmically.  Plots of all the $O-C$ curves can be found in Appendix A.  The signs of spot rotation frequency with respect to the binary orbit are not defined by the $O-C$ curves, but we summarize the prograde, retrograde, and `others' categories in Table \ref{percents}.  The blue vertical arrows mark the range of values of $k$ from Reinhold et al.~(2013) for single stars with $T_{\rm eff}$ in the range of 4800 K to 6500 K (derived from Eqn.~(\ref{eqn:kappa}) after multiplying by 0.37 d for the median period of our binaries; see Fig.~\ref{fig:pdist}).}
\label{diffRot}
\end{center}
\end{figure}

We utilized the fits of Eqn.~(\ref{autocorrFit}) to the autocorrelation functions computed for the $O-C$ curves of all of the 414 binaries in this study. From the B coefficient for each system, we evaluated a representative value for $P_{\rm mig}$. This value was then used to compute $|\kappa|$. Some 390 of the 414 systems yielded meaningfully measured values for $\kappa$; the distribution of $|\kappa|$ for this sample is shown in Fig.~\ref{diffRot}. We find that most systems have values of $|\kappa|$ that lie between 0.0003 and 0.005, but with another $\sim$20\% having values out to $|\kappa| \simeq 0.03$.

\begin{figure}
\begin{center}
\includegraphics[width=\columnwidth]{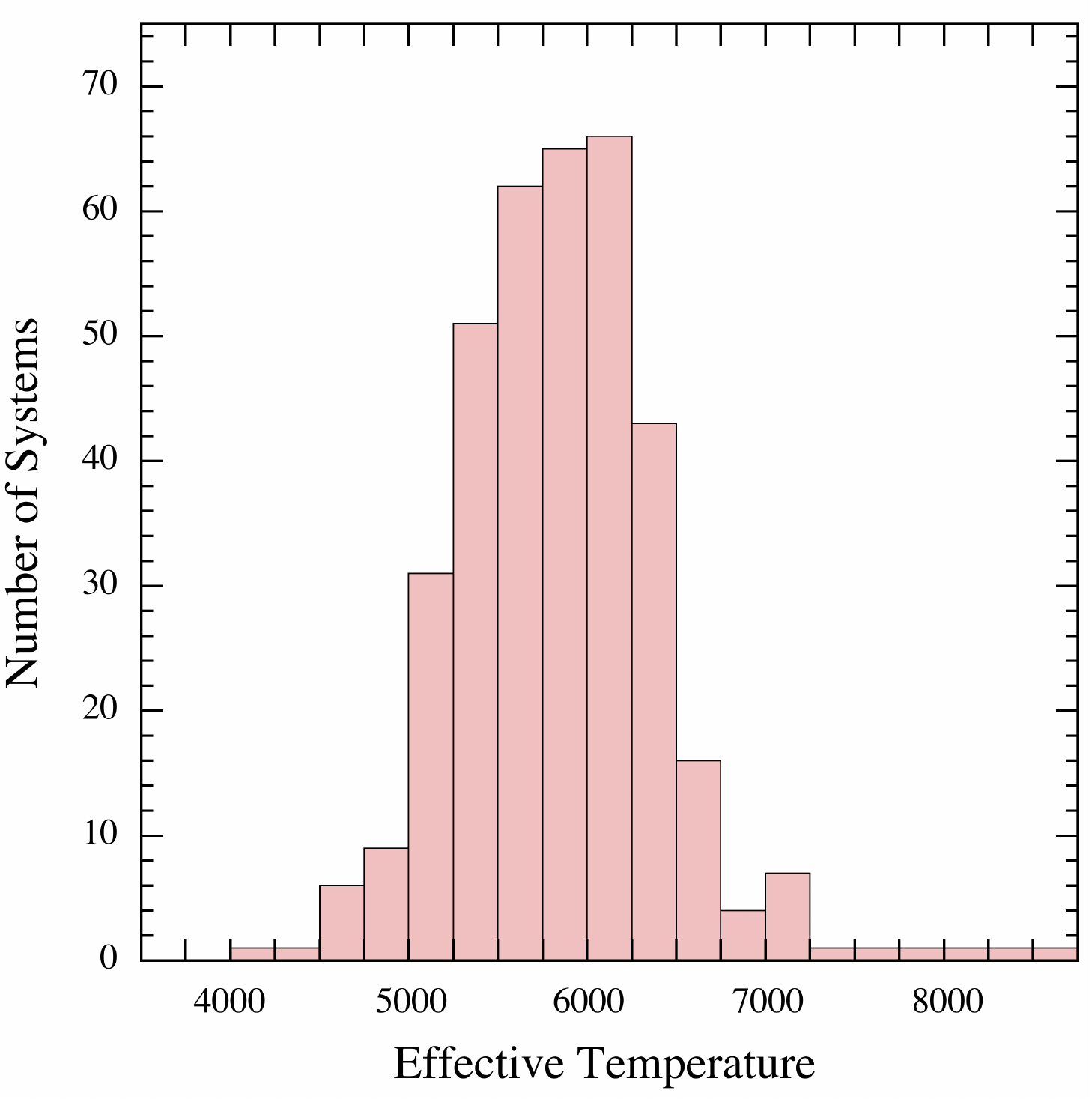}
\caption{A histogram of the effective (i.e., composite) temperature ($T_{\rm eff}$ taken from the KIC) for each of 390 contact binaries with anticorrelated primary and secondary $O-C$ curves that yielded measurable values of $\kappa$.  The vast majority of the systems are characterized by a value of $T_{\rm eff}$ in the range 5000 to 6500 K.} 
\label{Teff}
\end{center}
\end{figure}

Reinhold et al.~(2013) have carried out an
extensive study of differential rotation in 40,660 active {\em Kepler}
stars\footnote{These stars are mostly either single or in non-interacting binaries.  Moreover,
we note that the rotational periods covered in that study are limited to $P_{\rm rot} \gtrsim 0.5$
days, so not quite as short as the binary periods we are investigating.}.  
In some 18,600 of these stars they find two or more close
rotation periods which they take as evidence for differential
rotation and which they then use to derive limits on the differential rotation 
properties. Their Fig.~15 summarizes the measured values of horizontal shear
differential rotation parameter $\Delta \Omega \equiv \Omega_{\rm eq}-\Omega$, which is defined as
the difference in rotation frequency between the equator and the pole.
For stars hotter than $\sim$6400 K, values of $\Delta \Omega$ are
often found to exceed 0.2 rad d$^{-1}$.  However, for stars with $T_{\rm eff}$ 
in the range of 4800 K to 6400 K, which covers most of the range we are 
studying here (see Fig.~\ref{Teff}), $\Delta \Omega$ 
spans values of $\sim$0.035 to $\sim$0.2 rad d$^{-1}$.  This translates
to our fractional differential rotation parameter, $\kappa$ as:
\begin{equation}
\label{eqn:kappa}
\kappa \equiv \frac{\Delta \Omega}{\Omega_{\rm eq}} \simeq 0.0055 \rightarrow 0.03 \times P({\rm days})~,
\end{equation} 
where $P=2\pi/\Omega_{\rm eq}$.  
The range given by Eqn.~(\ref{eqn:kappa}), after multiplying
by the median orbital period in our sample of 0.37 days (see Fig.~\ref{fig:pdist}), 
is $\kappa \simeq$ 0.002 $\rightarrow$ 0.01,  
well within the range of values shown in Fig.~\ref{diffRot}, though 
centered on a value that is about a factor of three higher than the mean of the 
histogram in Fig.~\ref{diffRot}.  This modest difference might lead to the
speculation that the differential rotation of many of the stars in tight binaries 
is less pronounced than that of single stars or wide binary systems.

\begin{figure*}
\begin{center}
\includegraphics[width=0.9\textwidth]{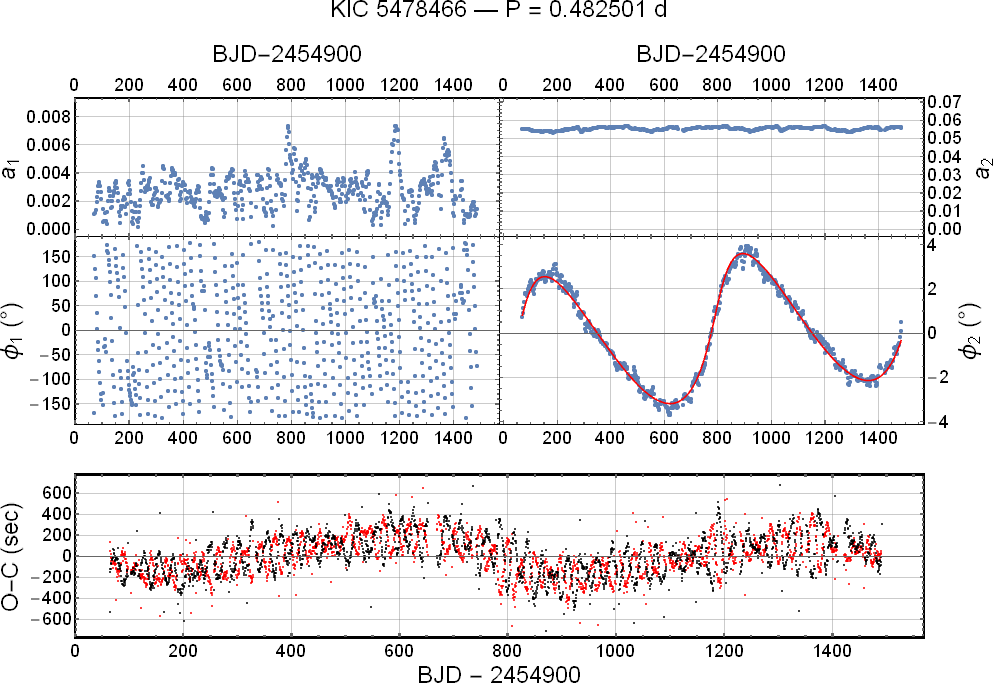}
\caption{The fitted parameters for KIC 5478466, as well as its $O-C$ curve, plotted as a function of time. The $\phi_2$ curve has the characteristic shape (but of opposite sign and twice the amplitude) of an $O-C$ curve dominated by a classical Roemer delay induced by a third star in the system.  The superimposed plot (red) is a best-fit eccentric Keplerian orbit.}
\label{tripleFigure}
\end{center}
\end{figure*}

In direct regard to close binary systems, Hill et al.~(2014) have measured the star spot rotation on the donor star in the cataclysmic variable star AE Aqr with an orbital period of 0.411 days.  The observations were made via Roche tomography of the K4V donor star in the system.  They find from two observations separated by 9 days, that there are two distinct bands of spots (at latitudes of 22$^\circ$ and 43$^\circ$) that have `lap times' (referred to in our work as $P_{\rm mig}$) of 262 and 269 days.  This corresponds to $|\kappa| \sim$0.0016, quite near the center of the distribution in Fig.~\ref{diffRot} before taking into account the $\cos^2 \alpha$ factor, but toward the lower end of the distribution after taking it into account.  By contrast, Hussain et al.~(2006) showed that the pre-cataclysmic variable star V471 Tau, with an orbital period of 0.521 days, has no measurable $\Delta \Omega$ with a constraint of $\lesssim 6$ mrad d$^{-1}$.  This translates according to Eqn.~(\ref{eqn:kappa}) to $\kappa \lesssim 5 \times 10^{-4}$.  Such a value corresponds to the lower limit on what we are able to measure in the current work.  However, these two examples can serve to indicate the importance of measuring a large sample of close binaries before drawing any sweeping conclusions.

\section{Finding Triples with $\phi_2$}
\label{findingTriples}

We noted earlier that we expect $\phi_2$ to remain constant {\em assuming no external perturbations} to the binary orbit. Therefore a systematically varying $\phi_2$ can allow us to infer the presence of an external perturbation to the binary, such as a third body. In Fig.~\ref{tripleFigure}, we present the example of KIC 5478466, where the \emph{correlated} behavior in its $O-C$ curve is characteristic of a binary that is actually part of a triple system (see Conroy et al.~2013). Note that in Fig.~\ref{tripleFigure}, $\phi_2$ exhibits the same variation, but with the opposite sign as that in the $O-C$ curve (as can be shown formally from the respective definitions of $O-C$ and $\phi_1$). The parameters we find for this triple are the period of the third body's orbit $P_{\rm trip} = 740.5 \pm 2$ days ($763 \pm 72$ days); the eccentricity of its orbit $e = 0.499 \pm 0.03$ ($0.459 \pm 0.003$), and the projected semimajor axis of the CM of the binary system in its orbit about the third star $a_{\rm bin} \sin i = 210.7 \pm 1.5$ sec ($206 \pm 13$ sec), where the values in parentheses are those reported by Conroy et al.~(2013). Thus it seems that we can also use the $\phi_2$ curves to search for new triple star systems (see Rappaport et al.~2013), without having the anticorrelated behavior of the $O-C$ curves making the correlated behavior more difficult to discern.

In this way, we were able to identify 39 triple systems where we infer the presence of a third body in the system via Roemer delays inscribed on the $\phi_2$ curves. All but four of these had been previously reported by Rappaport et al.~(2013) and Conroy et al.~(2013). 

\section{Summary and Conclusions}

Tran et al.~(2013) presented a spot model involving stellar differential rotation to explain anticorrelated oscillations in the $O-C$ curves for the times of primary and secondary minima in 32 short-period {\em Kepler} binaries. In this work, we used similar assumptions to those in that model to track explicitly the movement of spots around their host stars for 414 very short-period {\em Kepler} binaries with anticorrelated $O-C$ curves (see Appendix A and Fig.~\ref{diffRot}).   The success of our phasing analysis further validates the model involving spots that do not quite corotate with the binary orbit.

We found that about $\sim$34\% of the binaries exhibited spots moving in a retrograde sense with respect to the binary orbit, while $\sim$13\% exhibited spots moving in a prograde manner. The remaining binaries had either erratically moving spots---which could rather be caused by the appearance or disappearance of spots---or spots that did not move much in stellar longitude (see also Tran et al.~2013).

There are { at least three plausible causes} that could result in this spot motion relative to the binary. The first is that the entire body of the spotted star could be rotating asynchronously with respect to the orbit (i.e., not tidally locked stars). The second is that the equatorial zone, for example, of the star is synchronized, but that higher latitudes are rotating at different rates. In the latter case, individual stars (at the very least those that carry a spot) may be rotating more slowly (34\%) or more rapidly (13\%) with increasing stellar latitude.  { A third possibility is that the spot can migrate in longitude due a physical effect such as a systematically time-varying magnetic field configuration that might drive the location where spots form.} The observations could, of course, also be explained by a combination of asynchronously rotating stars, differential rotation in individual stars, and spot migration.  Finally, we note that the {\em jitter} that exists in nearly all of our $\phi_1$ curves is too large to be due solely to changes in the angular velocity of the stars (due to insufficient torques on the star), so asynchronous stellar rotation cannot be a sufficient explanation. 

Assuming that the explanation for the moving spots is dominated by differential stellar rotation, we can estimate the distribution of differential rotation values for 390 of the 414 binaries in this study. We actually computed the distribution of a parameter $|\kappa|$ which is formally defined as $P_{\rm orb}/P_{\rm mig}$ (see Eqn.~\ref{abskappa}) that can be related to the more standard differential rotation constant by $\kappa = k \cos^2 \alpha$, where $\alpha$ is the spot colatitude. We showed that since $\cos^2 \alpha$ is likely within a factor of 2 of unity in the systems of interest, $|\kappa|$ can serve as a meaningful proxy for $k$. We find that $|\kappa|$ mostly ranges between 0.0003 and 0.005, but a modest fraction ($\sim$20\%) of the fitted values range up to $\kappa \simeq 0.03$.  The values of $k$ found for single stars and non-interacting binaries (Reinhold et al.~2013; Eqn.~\ref{eqn:kappa}) are solidly in this range of values of $|\kappa|$ after multiplying in Eqn.~(\ref{eqn:kappa}) by 0.37 days which is the median orbital period of our sample.  The Reinhold et al.~(2013) values of $k$ are then only modestly shifted toward the higher end of our $|\kappa|$ distribution, by a factor of $\sim$3 (Fig.~\ref{diffRot}).  This suggests two things.  First, the rotation of {\em observed} spots in our sample of close binaries { most likely} is {\em not generally completely synchronized} with the orbital motion.  Second, the difference between the spot rotation rate and the orbital rate is often smaller, i.e., by a factor of $\sim$3 on average, than would be typical of simple differential stellar rotation in isolated stars. 

Another byproduct of our analysis is a new method of looking for third bodies among contact binaries. Just as one would search for correlated behavior in $O-C$ curves that would be indicative of a third body, one can use $\phi_2$ curves like those that we generated for our 414 contact binaries. With the $\phi_2$ curves, though, there is the additional benefit of not having anticorrelated effects that { can} obscure the relevant, long-term correlated effects, which often have a similar amplitude scale.

\acknowledgments
We thank Tam\'as Borkovits, Katalin Olah, and Slavek Rucinski for very useful discussions, { and an anonymous referee who made a number of helpful suggestions -- including the possibility of spot migration due to time-varying magnetic field configurations}. The authors are grateful to the {\em Kepler} Eclipsing Binary Team for generating the catalog of eclipsing binaries utilized in this work.  We also thank Roberto Sanchis-Ojeda for his help in preparing the stitched data sets.  B.C.'s work was performed under a contract with the California Institute of Technology funded by NASA through the Sagan Fellowship Program.


\appendix

\section{$O-C$ Curves for 414 {\em Kepler} Binaries With Anticorrelated Primary and Secondary ETVs }
\label{app:appA}

We show here the $O-C$ curves for both the primary and secondary minima of 414 short-period {\em Kepler} binaries.  These are systems that were chosen because the primary and secondary $O-C$ curves appear to be partially or totally anticorrelated.  In all there are 23 plots each displaying the $O-C$ curves for 18 binaries, in a $3 \times 6$ panel arrangement.  The first of the 23 plots is shown here, while the remaining 22 are contained in the online-only version of the Journal.

Because the $O-C$ curves were produced with a code different from 
that used by Tran et al.~(2013), we provide here a brief explanation 
of the calculations performed to obtain $O - C$ times for the primary 
minima from the detrended {\em Kepler} LC light curves.  

Detrending of each light curve began by applying a boxcar filter with
a duration of one orbital period (rounded to an integer number of long
cadence time intervals).  The filtered light curve was subtracted from the
unfiltered curve and the overall mean of the unfiltered curve was added 
to the differences to form the detrended light curve.  The latter is,
essentially, a high-pass filtered version of the original light curve.
In the following parts of this section, references to ``light curve"
shall be taken to mean the detrended light curve.

The basis of our timing method is the comparison of a selected
interval of a folded light curve with equal duration intervals of the
flux time series.  The folded light curve is produced by accumulating
the flux from each measurement in a histogram, where the bins of the
histogram are $\tau_{\rm LC} = 1765.46$ s in duration to match the
duration of the bins in the flux time series.  The bin assignment is
computed from:
\begin{equation}
\label{eqn:bin}
{\rm bin} = {\rm Int}\left\{ \frac{\left(t_{\rm obs}-t_0\right)}{\tau_{\rm LC}} -{\rm Int} \left[ \left(t_{\rm obs}-t_0\right)/P_{\rm orb} \right] \frac{P_{\rm orb}}{\tau_{\rm LC}}\ \right\}
\end{equation}
where `Int' is the integer function, $P_{\rm orb}$ is the best
estimate of the orbital period (taken either from the {\em Kepler}
binary catalog or our own determination), and $t_0$ is an arbitrary
epoch time.  The desired segment of the fold is typically of duration
30\% $\times P_{\rm orb}$ and is centered on the phase of the primary
minimum.  The comparison of the selected interval of the fold to an
equal length interval of the flux time series is accomplished via a
fit that adjusts a single scaling parameter $a$ -- defined by $lc(t) =
a \times fl(\phi)$ where $lc(t)$ is the light curve and $fl(\phi)$ is
the folded light curve -- to yield the minimum value of $\chi^2$ as a
function of the value of $a$.  The same segment of the folded light
curve is then used to fit {\em every} equal length segment of the unfolded
light curve.  The values of $\chi^2$ resulting from these fits form a
time series with the same cadence as the flux time series.  Each local
minimum in the values of $\chi^2$ considered as a time series is
identified with an $O-C$ time with an accuracy of $\pm \tau_{\rm
LC}/2$.

The {\em Kepler} data allow the determination of $O-C$ times with
accuracies that are much better than half of the long cadence bin time
of the data. Indeed, we and others have achieved accuracies of about
10 seconds in the best cases.  Some form of interpolation is necessary
to achieve such accuracies, but simple interpolation algorithms are
often not sufficient for obtaining times with accuracies better than
$\sim$100 seconds.

Our interpolation procedure is run for each coarsely determined $O-C$
time that was determined as described above.  In this interpolation procedure, 
the fit of the given segment of the flux time series is refit with each of a set of
folded light curves.  The values of $t_0$ of these folds are uniformly
spaced over the range of $\pm \tau_{\rm LC}/2$ with respect to the
value used in the initial fit.  Each fit is done over the phase
interval used in the initial fit.  Typically the fitting is done for
10 to 20 values of $t_0$.  The best value of $t_0$ is obtained from
the fit with the minimum value of $\chi^2$. The supposition is that this
fit corresponds to the alignment that results in the best match of the
minimum value in the flux time series with the local minimum in the
folded light curve; varying $t_0$ in this way acts like a `vernier
scale' on the phase measurement.  Simple interpolation using the
$\chi^2$ values for the immediate neighboring values of $t_0$ yields
the final value for this $O-C$ time.

$O - C$ times for the secondary minima were obtained in a similar
fashion, except that for each {\em Kepler} target we searched for
local minima roughly midway between primary minima.  This does not
work well, needless to say, for the secondary minimum of a highly
eccentric orbit.
 
\begin{center}
\begin{deluxetable}{lc|lc|lc|lc|lc|lc}
\tablewidth{0pt}
\tabletypesize{\tiny}
\tablecaption{\label{tab:Period} {\em Kepler} Binaries Exhibiting Anticorrelated $O-C$ Curves
}
\tablehead{
	\colhead{KIC \#} &
	\colhead{Period} &
	\colhead{KIC \#}  & 
	\colhead{Period} &
	\colhead{KIC \#} &
	\colhead{Period} &
	\colhead{KIC \#} &
	\colhead{Period}  &	
	\colhead{KIC \#}  & 
	\colhead{Period} &
	\colhead{KIC \#} &
	\colhead{Period}  \\
	\colhead{{\em Kep.} ID} & 
	\colhead{(days)} &
	\colhead{{\em Kep.} ID} & 
         \colhead{(days)} &
         	\colhead{{\em Kep.} ID} &
	\colhead{(days)} & 
         \colhead{{\em Kep.} ID} &
	\colhead{(days)} &
	\colhead{{\em Kep.} ID} &
         \colhead{(days)} &
	\colhead{{\em Kep.} ID} &
	\colhead{(days)}\\
\null{} \vspace{-0.2cm}}
\startdata 
1572353& 0.228931&  5005570& 0.422844&  6470998& 0.323154&  7884426& 0.296732&  9283826& 0.356524& 10724533& 0.745092\\ 
 1873918& 0.332433&  5008287& 0.291877&  6471048& 0.297228&  7889628& 0.342396&  9290838& 0.363032& 10794878& 0.432771\\ 
 2017803& 0.305742&  5022573& 0.441724&  6504095& 0.306775&  7936219& 0.385598&  9345838& 1.045874& 10797126& 0.432440\\ 
 2141697& 0.331042&  5033682& 0.379916&  6507541& 0.327212&  7941050& 0.297981&  9347868& 0.318470& 10799558& 0.366479\\ 
 2159783& 0.373884&  5077994& 0.696287&  6521123$^\dag$ & 0.366393&  7953796& 0.317548&  9347955$^\dag$ & 0.767372& 10801703& 0.313092\\ 
 2302092& 0.294673&  5123176& 0.707844&  6529902$^\dag$ & 0.284796&  7973882$^\dag$ & 0.346633&  9389122& 0.349950& 10809970& 0.447055\\ 
 2305277& 0.378772&  5166136& 0.233948&  6549491& 0.764654&  7985167& 1.384478&  9414905& 0.456526& 10847118$^\dag$ & 0.293553\\ 
 2437038& 0.267678&  5195137& 0.324117&  6593575& 0.345744&  8004839& 0.343764&  9419603& 1.052417& 10854621& 0.375939\\ 
 2447893& 0.661620&  5215999& 2.531158&  6615041& 0.340085&  8039225& 1.794127&  9451598& 0.362349& 10877703& 0.436937\\ 
 2570289& 0.279028&  5218385& 0.281538&  6671698& 0.471525&  8053107& 0.388871&  9466316& 0.352535& 10924462$^\dag$ & 0.375373\\ 
 2571439& 0.320979&  5283266& 0.314649&  6692340& 0.674849&  8108785& 0.228827&  9469350$^\dag$ & 0.397096& 10965091& 0.235856\\ 
 2694741& 0.326647&  5283839& 0.315231&  6766325& 0.439966&  8111387& 1.273597&  9470175& 0.303832& 11012801& 0.367311\\ 
 2695740& 3.615908&  5300878& 1.279443&  6778050& 0.945828&  8143757& 0.356523&  9481674& 0.388860& 11013608$^\dag$ & 0.318287\\ 
 2715007& 0.297111&  5307780& 0.308851&  6803335& 1.110853&  8183850& 0.398359&  9508052& 0.279996& 11017077& 0.708564\\ 
 2715417& 0.236440&  5371109& 1.191862&  6836140& 0.487721&  8190491& 0.777877&  9519590& 0.330894& 11036301& 0.285237\\ 
 2717141& 0.394571&  5374883& 0.419717&  6843507& 0.343182&  8190613& 0.332585&  9532591& 0.358375& 11083888& 0.342667\\ 
 2717423$^\dag$ & 0.398413&  5374999& 0.457880&  6863840& 3.852734&  8222945& 0.451123&  9639265& 0.506353& 11084647& 0.311674\\ 
 2835289$^\dag$ & 0.857762&  5377349& 1.116766&  6880727& 0.412737&  8230815& 0.345219&  9639491& 0.344546& 11084782& 0.586535\\ 
 2854432& 0.322326&  5389616& 0.406884&  6962901$^\dag$ & 0.977167&  8241252& 0.354044&  9649493& 0.371569& 11091082& 0.385082\\ 
 2858322& 0.436400&  5390913$^\dag$ & 1.680219&  6964796& 0.399966&  8248812& 0.685366&  9654476& 0.396324& 11127048& 0.299323\\ 
 3120742$^\dag$ & 0.236206&  5425950& 0.378168&  6966265& 0.313435&  8280135& 0.286885&  9658832$^\dag$ & 0.456847& 11137143& 0.313533\\ 
 3218683& 0.771670&  5426665$^\dag$ & 0.389553&  7023917& 0.772842&  8285349& 0.667110&  9700154& 0.257253& 11151970& 0.311629\\ 
 3232823& 0.423907&  5440746& 0.482654&  7035139& 0.309741&  8298344& 0.302875&  9702641& 0.503178& 11153627& 0.561697\\ 
 3339563& 0.841231&  5444392& 1.519528&  7117153& 0.359529&  8309175& 0.299591&  9713664& 0.321236& 11190836& 0.404016\\ 
 3342425& 0.393415&  5450322& 0.424019&  7118621& 0.370553&  8452840& 1.201160&  9772642& 0.275432& 11193447& 0.625884\\ 
 3431321& 1.015049&  5468295& 0.352012&  7118656& 0.321355&  8479107& 0.767577&  9786798& 0.445468& 11198068& 0.400175\\ 
 3437800& 0.362679&  5478466& 0.482501&  7119757& 0.742922&  8481574& 0.326865&  9821923& 0.349533& 11244501& 0.296676\\ 
 3443519& 0.353627&  5480282$^\dag$ & 0.360973&  7119876& 0.366364&  8545456& 0.315205&  9832227& 0.457948& 11245381& 0.306992\\ 
 3448245& 0.513490&  5534204& 0.275099&  7130044& 0.297665&  8548416& 1.163648&  9838047& 0.436161& 11284547& 0.262440\\ 
 3454864& 0.294120&  5534702& 1.025466&  7173910& 0.402244&  8555573& 0.326362&  9874575$^\dag$ & 0.335568& 11295026& 0.290855\\ 
 3545476& 0.317689&  5557368& 0.297931&  7176440& 0.358271&  8579707& 0.788592&  9934052& 0.352246& 11303416& 0.320784\\ 
 3557421& 0.393728&  5559529$^\dag$ & 0.376326&  7199353& 0.281987&  8587792& 0.368169&  9944907& 0.613440& 11305087& 0.309267\\ 
 3732732& 0.396187&  5563814& 0.308234&  7204041& 0.331679&  8590527& 0.739770&  9945280& 1.303460& 11336707& 0.251298\\ 
 3743834& 0.273615&  5567438& 0.584540&  7217866& 0.407157&  8608490& 1.082810&  9948201& 0.303181& 11402381& 0.417365\\ 
 3745184& 0.304233&  5620981& 0.333516&  7259917& 0.384679&  8652599& 1.360993&  9956124& 0.362725& 11404758& 0.351246\\ 
 3756730& 0.379164&  5699617& 0.288917&  7269797& 0.281792&  8682849& 0.352555&  9957411& 0.374690& 11405559& 0.284937\\ 
 3832382& 0.272649&  5783676& 0.401445&  7272739& 0.281165&  8690104& 0.408774&  9991887& 0.322396& 11442348& 0.312841\\ 
 3833859& 0.431743&  5790912& 0.383317&  7280704& 0.324350&  8715667& 0.405708&  9995660& 0.418590& 11460346& 0.385648\\ 
 3837677& 0.461984&  5802834& 1.092443&  7282128& 0.347050&  8823666& 0.432451&  9995981& 0.293267& 11494583& 0.248342\\ 
 3848042& 0.411453&  5809868& 0.439390&  7284688& 0.646043&  8846978& 1.379060& 10032392& 0.391157& 11496078& 0.299719\\ 
 3853259& 0.276648&  5821050& 1.933369&  7304911& 0.625551&  8872737& 0.459099& 10119517& 0.737494& 11498689$^\dag$ & 0.306307\\ 
 3936357& 0.369154&  5823121& 2.298001&  7339123& 0.347103&  8878719& 0.364635& 10123627& 0.294939& 11502581& 0.333636\\ 
 3956545& 0.330280&  5880661$^\dag$ & 0.330287&  7348206& 0.401205&  8934111& 0.364518& 10128961& 0.347743& 11559864& 0.383301\\ 
 3956881& 0.696946&  5951553& 0.431971&  7368451& 0.373377&  8956957& 0.324382& 10135584& 0.391294& 11566174& 0.276875\\ 
 3972629& 0.244240&  5955731& 1.003830&  7506446& 0.352600&  8982514& 0.414491& 10148799& 0.346605& 11572643& 0.345404\\ 
 4056791$^\dag$ & 0.361263&  5961327& 0.299202&  7512381& 0.423925&  9002076& 0.481405& 10154189& 0.411239& 11672319& 0.331775\\ 
 4244929& 0.341403&  5984000& 0.323058&  7516345& 0.491870&  9004380& 0.250333& 10155563& 0.360267& 11704155& 0.273867\\ 
 4249218& 0.316195&  6024572& 0.414910&  7518816& 0.466580&  9020289& 0.384028& 10253421& 0.428354& 11716688& 0.301218\\ 
 4451148& 0.735981&  6029214& 0.819066&  7523960$^\dag$ & 0.326296&  9020426& 0.913393& 10255110& 0.336703& 11717798& 0.374715\\ 
 4464999& 0.434164&  6045264& 0.909310&  7529123& 0.350078&  9021397& 0.426211& 10257903& 0.858570& 11751847& 0.331484\\ 
 4474193& 0.322505&  6050116& 0.239908&  7533807& 0.316845&  9030509& 0.401785& 10264202& 1.035148& 11805235& 0.395193\\ 
 4569923& 0.313583&  6057829& 0.483427&  7542091& 0.390499&  9032671& 0.249608& 10280990& 0.384457& 11873166& 1.468852\\ 
 4661397& 0.292325&  6061139& 0.322441&  7546791& 0.242361&  9071104& 0.385213& 10288502& 0.230185& 11910076& 0.348123\\ 
 4669592& 0.378428&  6067735& 0.446181&  7584826& 0.622309&  9071373& 0.421769& 10291683& 2.043743& 11920266& 0.467410\\ 
 4672934& 0.382747&  6118779& 0.364246&  7597095& 0.301206&  9087918& 0.445607& 10322296& 0.315033& 12019674& 0.354504\\ 
 4677321& 1.572181&  6128248& 0.291744&  7610486& 0.457679&  9091810& 0.479722& 10322582& 0.291269& 12055421& 0.385606\\ 
 4758092& 0.289166&  6143826& 1.354760&  7625150$^\dag$ & 0.394215&  9094335& 0.340759& 10350225& 0.282070& 12066630& 1.098374\\ 
 4762887& 0.736574&  6264091& 0.325018&  7660607$^\dag$ & 2.763401&  9097798& 0.334065& 10351767& 0.635070& 12071741& 0.314264\\ 
 4843222& 0.387405&  6265720& 0.312428&  7676610& 1.225675&  9114111& 0.381261& 10355671& 0.303630& 12104285& 0.251451\\ 
 4850874& 1.775901&  6281103& 0.363281&  7680593& 0.276391&  9142964& 0.589253& 10357852& 0.488341& 12108333& 0.705448\\ 
 4904304& 0.389474&  6286402& 0.343320&  7691553& 0.348308&  9145707& 0.320771& 10388897& 0.343729& 12109575& 0.531655\\ 
 4909422& 0.395103&  6353203& 0.508504&  7700181& 0.363699&  9145846& 0.294480& 10389809& 0.397397& 12153332& 0.294627\\ 
 4921906& 0.213732&  6358198$^\dag$ & 2.623527&  7709086& 0.409475&  9151972& 0.386796& 10447902& 0.337464& 12157987& 0.577844\\ 
 4936334& 0.354369&  6359771& 0.406068&  7744173& 0.398528&  9178425& 0.518667& 10485137& 0.445275& 12305537& 0.361815\\ 
 4937350$^\dag$ & 0.393664&  6368316& 0.338505&  7773380& 0.307578&  9179806& 0.280963& 10528299& 0.399802& 12356746& 1.004908\\ 
 4941060& 0.303422&  6424124& 0.385528&  7818448& 0.618952&  9237755& 0.331670& 10600319& 0.281516& 12418274& 0.352722\\ 
 4945588& 1.129080&  6431545& 0.316834&  7821450& 0.314762&  9245929& 0.308375& 10680475& 0.349832& 12418816& 1.521870\\ 
 4950505& 0.294809&  6443392& 0.776976&  7839027& 0.720693&  9268105& 0.425694& 10711938& 0.357604& 12458133& 0.332522\\ 
 4991959& 0.360939&  6467389& 0.288976&  7877062& 0.303651&  9274472& 0.302091& 10723143& 0.307642& 12598713& 0.257179\\ 
\enddata
\null{} \vspace{-0.2cm}
\tablecomments{414 {\em Kepler} targets exhibiting anticorrelated $O-C$ curves for their primary and secondary eclipses.  The {\em Kepler} targets marked with a $\dag$ are listed as `false positives' in the {\em Kepler} binary star catalog.  Their anti-correlated $O-C$ curves, however, may well indicate that they are, in fact, true short period binaries.}
\end{deluxetable}
\end{center}

\newpage

\begin{figure}
\begin{center}
\includegraphics[width=0.95\columnwidth]{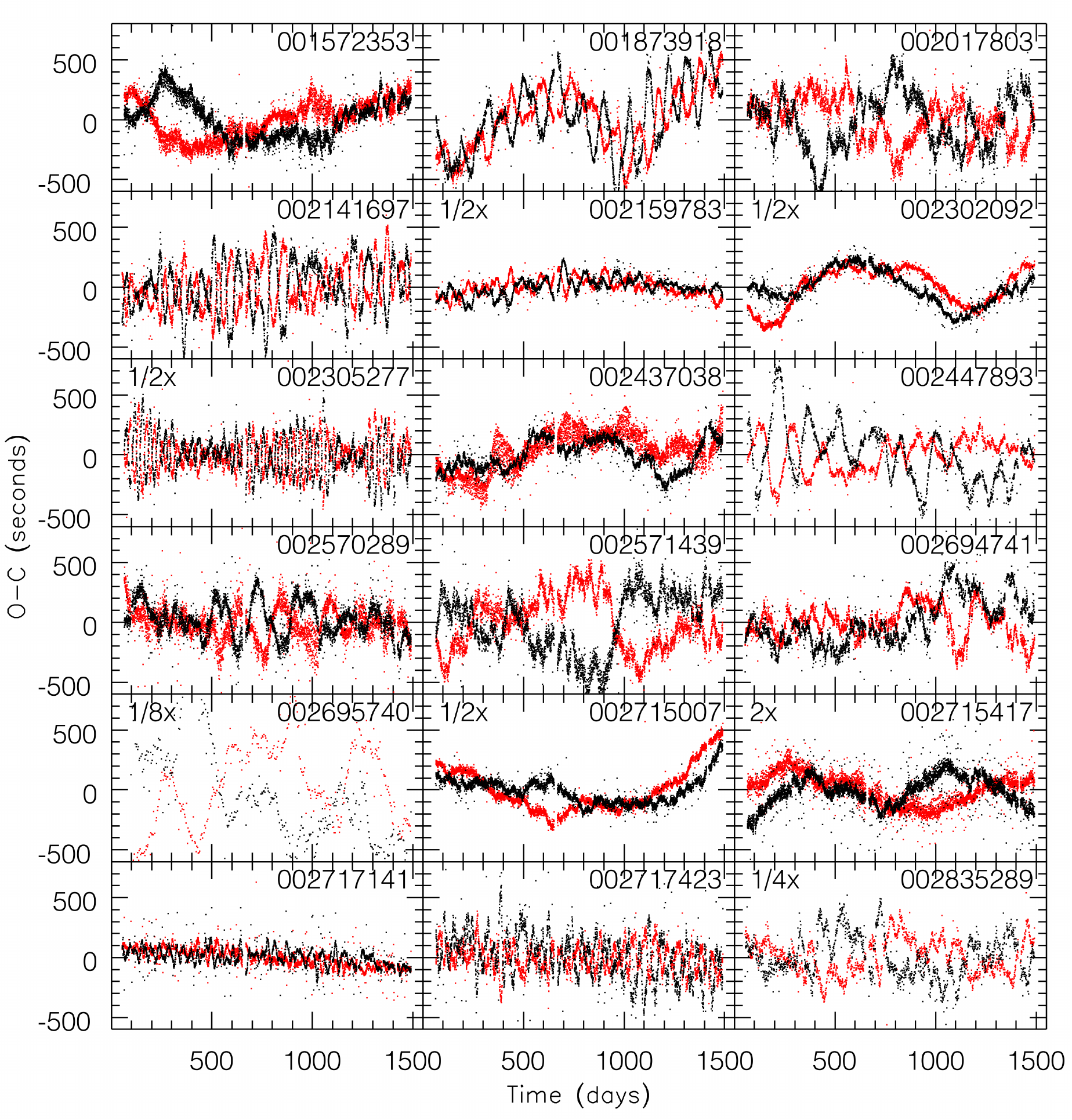}
\caption{A set of $O-C$ curves for 18 of the 414 short-period binaries that we studied in this work.  The $O-C$ curves for the primary are in black, while those for the secondary are in red.  The KIC numbers are written in the upper right corner of each panel.  The fraction, followed by a ``$\times$'' symbol, in the upper left corner of some panels is the amount by which the $O-C$ values have been {\em divided} for better visibility (e.g., $1/8 \times$ implies that one should multiply the displayed vertical scale by 1/8 in order to obtain the actual $O-C$ value). Figures 15 through 36, covering all the 414 binaries in this study are available in the online-only version of the Journal.}
\label{fig13}
\end{center}
\end{figure}

\end{document}